\documentclass[manuscript]{aastex}

\shorttitle{Study of globular clusters in NGC 5128}
\shortauthors{Asis Kumar Chattopadhyay et al.}

\begin{document}

\title{Study of NGC 5128 Globular Clusters Under Multivariate Statistical Paradigm}

\author{Asis Kumar Chattopadhyay\altaffilmark{1}}
\affil{Department of Statistics, Calcutta University, 35 B.C.
Road, Calcutta 700019, India} \email{akcstat@caluniv.ac.in}

\author{Tanuka Chattopadhyay\altaffilmark{2}}
\affil{Department of Applied Mathematics, Calcutta University, 92 A.P.C. Road, Calcutta 700009, India}
\email{tanuka@iucaa.ernet.in}

\author{Emmanuel Davoust\altaffilmark{3}}
\affil{Laboratoire d'Astrophysique de Toulouse-Tarbes,
Universit\'e de Toulouse, CNRS, 14 Avenue Edouard Belin, 31400
Toulouse, France } \email{davoust@ast.obs-mip.fr}

\author{Saptarshi Mondal\altaffilmark{1}}
\affil{Department of Statistics, Calcutta University, 35 B.C.
Road, Calcutta 700019, India}

\and

\author{Margarita Sharina\altaffilmark{4,5}}
\affil{Special Astrophysical Observatory, Russian Academy of Sciences, N. Arkhyz, KCh R, 369167, Russia}
\affil{Isaac Newton Institute of Chile, SAO Branch}
\email{sme@sao.ru}

\begin{abstract}

An objective classification of the globular clusters of NGC 5128
has been carried out by using a model-based approach of cluster
analysis. The set of observable parameters includes structural
parameters, spectroscopically determined Lick indices and radial
velocities from the literature. The optimum set of parameters for
this type of analysis is selected through a modified technique of
Principal Component Analysis, which differs from the classical one
in the sense that it takes into consideration the effects of
outliers present in the data. Then a mixture model based approach
has been used to classify the globular clusters into groups. The
efficiency of the techniques used is tested through the
comparison of the misclassification probabilities with those
obtained using the K-means clustering technique. On the basis of
the above classification scheme three coherent groups of globular
clusters have been found. We propose that the clusters of one
group originated in the original cluster formation event that
coincided with the formation of the elliptical galaxy, and that
the clusters of the two other groups are of external origin, from
tidally stripped dwarf galaxies on random orbits around NGC 5128
for one group, and from an accreted spiral galaxy for the other.

\end{abstract}

\section{Introduction}

Globular Clusters (GCs) are touchstones of astrophysics. Their
study addresses many important issues ranging from stellar
evolution to the formation of galaxies and cosmology. However
their origin and formation history, which are obviously linked to
that of their parent galaxy, are still poorly understood.

Classical formation of galaxies can be divided into five major
categories: (i) the monolithic collapse model,(ii) the major
merger model, (iii) the multiphase dissipational collapse model
(iv) the dissipationless merger model and (v) accretion and in
situ hierarchical merging.

According to the monolithic collapse model an elliptical galaxy is
formed through the collapse of an isolated massive gas cloud at
high redshift (Larson 1975; Carlberg 1984; Arimato \& Yoshi
1987). In this model the color distribution of GCs is unimodal and
the rotation of GCs is produced by the tidal force from satellite
galaxies (Peebles 1969).  In the major merger model elliptical
galaxies are formed by the merger of two or more disk galaxies (Toomre
1977; Ashman \& Zepf 1992; Zepf et al. 2000). Younger GCs are
formed out of the shocked gas in the disk while blue GCs come from
the halos of the merging galaxies (Bekki et al. 2002). As a result
the color distribution is bimodal. In this scenario, the kinematic
properties of the GCs depend weakly on the orbital configuration
of the merging galaxies, but the metal-rich GCs are generally
located in the inner region of the galaxy, and the metal-poor ones
in the outer regions.

The multiphase dissipational collapse has been proposed by Forbes
et al. (1997). According to this model the GCs form in distinct
star formation episodes through dissipational collapse. In
addition there is tidal stripping of GCs from satellite dwarf
galaxies. Blue (metal-poor) GCs form in the initial phase and red
(metal-rich) GCs form from the enriched medium at a later epoch,
thus producing a bimodal color distribution of the GCs. This model
predicts that the system of blue GCs has no rotation and a high
velocity dispersion while the red GCs show some rotation depending
on the degree of dissipation.  C\^ot\'e et al. (1998) proposed a
model in which the GC color bimodality is due to the capture of
metal-poor GCs through merger or tidal stripping. The metal-rich
GCs are the initial population of GCs in the galaxy and are more
centrally concentrated than the captured GCs. The main difference
with the previous model is that no age difference is expected
between the blue and red GC populations. The very different
origins for the two populations imply rather different orbital
properties, in particular the metal-poor GCs should show a larger
velocity dispersion than the metal-rich ones, comparable in the
outer region to that of the neighboring galaxies.

From the above discussion it appears that there are kinematic
differences among the sub populations of GCs in different
galaxies. These differences can be used as an observational
constraint on the galaxy formation model. In the above studies the
GCs are classified as metal-rich and metal-poor on the basis of
the value of a single parameter [Fe/H] $>$ or $<$ -1 which is
subjective in nature and also inappropriate in a multivariate
setup. Concentrating on a single parameter means that one ignores
the joint effect of several parameters.

With the above objective in mind we have carried out a
multivariate analysis of extragalactic GCs. In this context, NGC
5128 is of interest because it is the nearest giant elliptical
galaxy whose large sample of GCs is amenable to
spectroscopic observations (Beasley et al. 2008) and whose
structural parameters have been derived by fitting models to surface
brightness profiles based on HST/ACS imaging (McLaughlin et al.
2008). Also the radial velocities are available for a large
subsample of GCs (Woodley et al. 2007).

In the present study we have first used a modified technique of
principal component analysis (PCA) (Salibi\'an-Barrera et al.
2006) to search for the optimum set of parameters which gives the
maximum variation for the GCs in NGC 5128. This method helps to
extract the significant parameters from the large set of
photometric, structural, and kinematic parameters. Then we have
classified the GCs on the basis of these significant parameters
using a model-based method of Cluster Analysis (CA) (Qui \&
Tamhane 2007) which finds the structure of the optimum groups of
GCs instead of choosing groups in an ad hoc manner on the basis of
a single parameter. This multivariate analysis helps to enunciate
a more efficient theory of GC formation.

In this context it should be mentioned that kinematic studies of
GCs in several giant elliptical galaxies show results which differ
from galaxy to galaxy (Woodley et al. 2007). In a recent study
(Hwang et al. 2008) the rotation of metal-poor and metal-rich GCs
have been studied in 6 giant elliptical galaxies, two systems of
GCs show strong rotation while the other ones show moderate or
weak rotation. We have studied NGC 5128 as the representative of
the latter group while it remains important to study
representatives of the former group in a multivariate setup, once
the adequate data are available.

In this paper the different data sets used are presented in
section 2. Section 3 gives a list of the different methods used in
the present study while the determination of spectroscopic ages
and metallicities is discussed in section 4. The results and
discussions are presented in section 5.  Brief discussions of the
methods used are given in the appendices.

\section{Data Set}

Our analysis is based on the sample of GCs of the early-type
central giant elliptical galaxy in the Centaurus group, NGC 5128,
whose structural parameters have been derived by fitting King and
Wilson models to the surface brightness profiles based on HST/ACS
imaging in the F606W bandpass (McLaughlin et al. 2008). The
distance is that adopted by McLaughlin et al. (2008), namely 3.8
Mpc. The sample consists of 130 GCs (3 outliers have been excluded
during cluster analysis) whose available structural and
photometric parameters are tidal radius ($R_{tid}$, in pc), core
radius ($R_{c}$, in pc), half-light radius ($r_{h}$, in pc),
central volume density (log$\rho_{0}$, in $M_{\odot}$ p$c^{-3}$),
predicted line of sight velocity dispersion at the
cluster center ($\sigma_{p,0}$ in km$s^{-1}$), two-body relaxation time at the
model projected half-mass radius ($t_{rh}$, in years),
galactocentric radius ($R_{gc}$, in kpc), concentration
(c$\sim$ log($R_{tid}$/$R_{c}$)), dimensionless central
potential of the best fitting model ($W_{0}$),
extinction-corrected central surface brightness in the F606W bandpass
($\mu_{0}$ in mag arcsec$^{-2}$), V surface brightness averaged
over $r_h$ ($<\mu_v>_h$) in mag arcsec$^{-2}$, integrated
model mass (log$M_{tot}$, in $M_{\odot}$), Washington $T_{1}$
magnitude, extinction corrected color $(C - T_{1})_{0}$ and
metallicity ([Fe/H],in dex) determined from the color $(C -
T_{1})_{0}$.

The radial velocities ($V_{r}$, in km$s^{-1}$) are available for
48 GCs (Woodley et al. 2007), the position angles ($\psi$, east of
north) were derived from the coordinates. There are
51 GCs in common with the sample of GCs observed by Beasley et al.
(2008) and the present sample. Among these 21
GCs have published Lick Indices (Beasley et al. 2008). These data
are used to derive the ages and metallicities ([Z/H]) of 21
GCs of our sample.

The entire data set of 130 GCs with all the parameters (from the
literature as well as derived by authors) are listed in Tables 1
and 2 together with their group membership as a result of CA.
The three outliers which have not been considered in the final
CA study are marked.

\section{Method}

In the present work we have used some statistical techniques
already developed for proper analysis of the data.

We have considered a robust principal component
analysis based on multivariate MM-estimators (Salibi\'an-Barrera
et al.2006). Principal Component Analysis is a very common
technique used in data reduction and interpretation in
multivariate analysis. The above mentioned method has been
developed to incorporate robustness property into the classical
PCA in order to estimate the effects of outliers present in the
data. In this method one MM-estimator of scatter is used
instead of sample covariances which are very much sensitive to
outliers. In particular the eigenvectors and eigenvalues of
multivariate MM-estimators of shape are used as introduced by
Tatsuoka \& Tyler (2000).

For Cluster Analysis we have used two methods : one is
based on Mixture Models and the other is a partitioning method.
The mixture model method provides a parametric approach to the
clustering problem proposed by Qui \& Tamhane (2007). Here the
Expectation-Maximization (EM) algorithm is used to compute the
maximum likelihood estimators (MLEs) of the parameters of the
model. These parameters include mixing proportions, which may be
thought of as the prior probabilities of different clusters; then
the maximum posterior (Bayes) rule for clustering has been used.

The partitioning method, known as K-means algorithm, is one
of the most popular method of clustering developed by
MacQueen(1967). This method is distribution-free in nature, but
cannot provide any estimate of misclassification error
probabilities of observations. Qui \& Tamhane (2007) proved that
the mixture model method is a better method of clustering since it
yields smaller expected misclassification rates. To find the
optimum number of clusters (i.e. the value of K) we have used the
method developed by Sugar \& James (2003).

In the present work we have also used the Levenberg-Marquardt
algorithm to compute the rotation amplitudes and position angles
of the axes of rotation of different groups obtained from cluster
analysis. They are listed in last two columns at the beginning of
each group of Table 2 as well as in Table 4.

All the above mentioned methods are discussed in brief in the
appendices.

\section{Determination of ages and metallicities}

We took advantage of the method of Lick indices (Faber 1973,
Worthey et al. 1994) to disentangle effects of age and metallicity
on integrated spectra of globular clusters. A three-dimensional
interpolation and $\chi^2$ minimization routine by Sharina,
Afanasiev \& Puzia (2006) (see also Sharina \& Davoust, 2009)
allowed us to estimate age, [Z/H], and alpha-element ratio for
each individual GC simultaneously. It minimizes the summed
difference over all Lick indices between
the observational and theoretical index values, weighed by the errors
of index measurements. The theoretical Lick indices were obtained
using linear interpolation on the grids of Simple Stellar
Population models of Thomas et al. (2003, 2004). The errors on
the evolutionary parameters depend on the errors of Lick indices
and on the accuracy of the radial velocities. The random errors
of Lick index measurement in individual spectra depend primarily
on the S/N ratio in the spectra. The typical source of systematic
errors of Lick indices is quality of calibrations of an
instrumental system into the Lick standard one (Worthey et al.
1994).

The comparison of our new metallicity determinations for the
entire data set of GCs from Beasley et al. (2008), based on their
published Lick indices, with metallicities from Beasley et al.
show a very good correlation (r $\simeq$ 0.9; Fig. 1). The
photometric
metallicities are available for all the GCs of our sample.\\

\section{Results and discussions}

\subsection{Analysis based on PCA}

In PCA our goal is to reduce the large number of parameters in a
data set to a minimum number while retaining a maximum variation
among the objects (here GCs) under consideration. The technique
therefore helps to sort out the optimum set of parameters that causes
the maximum overall variation in the nature of GCs in NGC 5128.
We initially excluded the observations
corresponding to C177 because the values of $R_{tid}$,$r_{h}$ and
$<\mu_v>_h$ for this GC are significantly higher
than those of all other GCs. We started with the
parameters $R_{tid}$, $R_{c}$, $r_{h}$, log$\rho_{0}$,
$\sigma_{p,0}$, $t_{rh}$, $R_{gc}$, c, $W_{0}$, $\mu_{0}$,
$<\mu_v>_h$, log$M_{tot}$, $T_{1}$, $(C - T_{1})_{0}$ and [Fe/H]
of 130 GCs and considered only a smaller set of parameters (selected
by trial and error of all possible combinations of
parameters). We determined the minimum number of principal
components on the basis of maximum percentage (90\%) of total
variation. Since the total set of all possible combinations is
very large only a few of the combinations of parameter sets
are given for the comparison in Table 3. We mention only
these combinations in Table 3 because for all the other combinations
the number of principal components to be preserved is higher.

Table 3 shows that sets S3, S4, S6 and S7 have a
minimum number of principal components (viz. 1). Among these S7
has the maximum variation corresponding to the first principal component. So
S7 has been selected as the optimum set. The parameters found in
S7 are the same as those used by Pasquato \& Bertin (2008) and
Djorgovski (1995) in constituting the fundamental plane (FP) for
half-light parameters of GCs in the Galaxy. The present set is
different from the FP found by McLaughlin (2000) and McLaughlin
\& van der Marel (2005) which includes luminosity (L), core mass-
to-light ratio ($\gamma_{V,0}$), binding energy ($E_b$) and
central concentration (c).

\subsection{Analysis based on Cluster Analysis}

Cluster analysis is the classification of objects into different
groups or more precisely the partitioning of a data set into
subsets (clusters) so that the data in subsets share some common
trait according to some distance measure.

In the previous section the most significant parameters were
filtered from a large number (here 15) of parameter sets through
the modified PCA which starts from the
matrix containing measures of shape parameters involved instead of
considering the correlation matrix as is generally done. The most
significant parameters responsible for the maximum variation, while
keeping the number of principal components minimum are
($<\mu_v>_h, r_h, \sigma_{p,0}$).

Next the cluster analysis based on the mixture model method is carried
out with respect to these three parameters. The optimum number of
groups is selected objectively by a widely used  method, K-means
clustering (MacQueen 1967), together with the method developed by
Sugar \& James (2003) for finding the optimum number of clusters.
The K-means method is necessary to find the optimum number of
groups which worked as input to the mixture model method. After
doing CA by K-means and associated optimum number method it is
found that optimality occurs at K=4 with only one GC, C156, in a
group. Then CA is performed again after removing this object and
with optimum criterion. Optimality then occurs at K=4 with again a
very small group containing two GCs, C169 and F1GC15. These GCs
are removed and the process is repeated with the sample of 127
objects. Now the optimum number of groups is found at K=3 with GCs
distributed into three evenly populated groups. We thus select
this sample for study and perform the new method of CA taking K =
3.

In order to establish the better performance of the model based
CA method, we have computed the expected
misclassification probabilities corresponding to some of the
parameters. In particular for $\sigma_{p,0}$ under the K-means
method it was found to be 0.4088 whereas under the new CA method
it is only 0.14978. For CA we have
removed three GCs which are outliers with respect to set S7 used
for CA.

The mean values (with standard errors) of all the parameters are
listed in Table 4. The rotation amplitudes, rotation axes,
projected velocity dispersions and rotation strength
($\Omega$R/$\sigma_{v}$ = x) of the three groups of GCs are also
listed. It is to be noted that for determining rotation amplitude,
rotation axis, projected velocity dispersion and rotation
strength, values of radial velocities ($v_{r}$) are needed (viz.
equations (C1) and (C2)). Since almost no radial velocities of
GCs in G3 are available in Woodley et al. (2007) for G3, the
rotation parameters were not derived for this group. The mean ages
and metallicities derived from spectra (Beasley et al. 2008)
using SSP models (Thomas et al. 2003, 2004) for these groups are
also included in Table 4. Ages and spectroscopic metallicities
are not available for the GCs in G3, which are rather faint.

\subsection{ Properties of globular clusters in three groups}

CA segregated the sample of GCs into three
groups, G1, G2, G3, according to their structural properties. We
emphasize that no property of the stellar populations, such as
spectroscopic element abundances, were taken into account.  Nor
did we use any information on the radial velocity of the GC
or on their position in the galaxy. Thus some of
the properties of the three groups discussed below are not a
consequence of this analysis.

The different properties of the three groups are presented in
Table 4. The main difference between the groups lies in the mean
luminosities ($T_1$) of the GCs and their individual central
velocity dispersions ($\sigma_{p,0}$), which are both indicators
of their individual masses. The GCs of G2 are the most massive,
followed by those of G1, and then G3.

It may be instructive to compare the structural properties of the
GCs in NGC 5128 and in our own Galaxy. Fig.2 shows that mass is
correlated with the other structural parameters c, $\mu_{0}$,
$\rho_{0}$, $\sigma_{p,0}$, like in the Galaxy (Djorgovski \&
Meylan, 1994). On the other hand, unlike in our Galaxy, ellipticity
(e), c, $\mu_{0}$  and $\rho_{0}$ are not correlated with $R_{gc}$.

The parameter $r_h$ is predicted to remain constant during the dynamical
evolution of GCs, it is thus interesting to compare its value in
our sample with that in other samples. We have compared the values
of $r_h$ (King model) with the $r_h$ measured in a large sample of
early-type galaxies by J\'{o}rdan et al. (2005). Note that these
authors estimated $r_h$ using a King rather than Wilson model.
Determining the peak of the distribution of $log(r_h)$ for our
whole sample with a nonparametric fit and Epanechnikov filter
(Epanechnikov 1969), and assuming that this peak should fall at
$r_h= 2.85\pm 0.3$ pc, we find a distance to NGC5128 of $3.58\pm
0.3$ Mpc. The distance adopted in section 2 is within the error
bars of this new distance estimate for NGC5128. The individual
groups are too small to show any clear peak in the distribution of
$r_h$.

G3 is different from the two other groups in that it is mainly
distributed in the outer regions of the galaxy (large $R_{gc}$,
see Fig.3). Because most GCs of G3 are of low mass and thus of
faint luminosity, no spectroscopy is available for them, and thus
no age, radial velocity or spectroscopically determined metal
abundance ([Z/H]). A possible consequence of this is that G3 might
be polluted by foreground stars, although McLaughlin et al. (2008)
only mention two possible such cases (C145, C152), by objects
resembling intermediate-age Galactic open clusters (van den
Bergh,2007), or by background galaxies.

The three groups have very different distributions of photometric
metallicity [Fe/H], as shown by their probability density
functions (PDF) in Figs. 4 and 5. The lines indicate
non-parametric density estimates using an Epanechnikov
kernel (Epanechnikov 1969). The bin width (0.2 dex for all groups)
was chosen using the whole sample and the Freedman-Diaconis rule
based on the sample size and the spread of the data (for a
definition, see $"Freedman-Diaconis-rule"$ in wikipedia. For an
explanation of the histogram as a density, see Freedman and
Diaconis (1981)).  A peak in the metallicity distribution at
$[Fe/H]\sim -1$ is seen in all three groups. The PDF of G3 is
clearly bimodal with a subgroup at very low metallicity, while the
metallicity range of the two other groups is more limited and more
in line with that of GCs in other galaxies, including our own. For
example, our Galaxy has a distribution of [Fe/H] which peaks at
-1.6 and -0.6 and the lowest value is -2.29 (Harris 2001). For
M31, M 81 and NGC 4472, the numbers are respectively : -1.4, -0.6
and -2.18 (Barmby et al. 2000) -1.45, -0.53 and -2.0 (Ma et al.
2005) -1.3, -0.1 and -2.0 (Geisler et al. 1996).

While metallicity and colors measure the state of evolution of the
stellar populations, the structural parameters mainly give
indications on the dynamical evolution of the GCs, and the three
groups are markedly different in this respect as well. GCs
are predicted to become
rounder as they lose stars and angular momentum in the course of
their evolution. In G1 the roundest GCs are also the most
metal-rich, whereas no such correlation is present in G2 or G3, as
shown on Fig.6. In G2, the roundest GCs are also the least
massive. During the dynamical evolution of GCs, their $R_c$
shrinks and their $R_{tid}$ increases. Core collapse occurs when
the ratio $log(R_{tid}/R_c) >$ 2.5. As shown on Fig.7, this is the
case for most GCs of G2 which are thus at an advanced stage of
dynamical evolution, for a minority of GCs in G1, and for hardly
any in G3.

A color-color diagram is another way of examining the properties
of the stellar populations of the GCs. This is done in Fig.8,
which shows $(C-T_1)_o$ vs $(M-T_1)_o$ in the Washington
photometric system for our sample, using data from Harris et al.
(2004). For comparison, we also plotted on Fig.8 data for Galactic
GCs (open squares, from Harris \& Canterna 1977) and for low
surface-brightness (LSB) dwarf galaxies (open triangles) from
Cellone et al. (1994).

In order to interpret this Figure, we also plotted several model
stellar populations : the thin black solid line is an SSP track of
varying metallicity at 15Gyr from Cellone \& Forte (1996). The
thick short-dashed and dotted lines are tracks of varying age for
ellipticals and Sa galaxies from Buzzoni (2005). The colored grid
of models (metallicities z=0.0004, 0.004, 0.008, 0.02, and 0.04)
are GALEV SSP models from Anders \& Fritze - v. Alvensleben,
(2003); they clearly show the extent of the age-metallicity
degeneracy in this diagram. The three sets of models are based on
different assumptions, and their differences are indicative of the
uncertainties involved. The reddening line is roughly parallel to
the tracks.

Before discussing Fig.8, we recall some relevant characteristics
of the Washington photometric system. The C band includes the
U-band and half of the B-band of the Johnson-Cousins photometric
system (e.g. Lejeune 1996), and is thus sensitive to the presence
of different features in the blue part of the color-magnitude
diagram, such as extreme horizontal branch stars and abnormally
wide main sequences, including branches of different colors. Such
features are characteristics of the most massive Galactic GCs,
probable cores of disrupted dwarf spheroidal satellites
(Recio-Blanco et al. 2006). $C-T_1$ is almost twice as sensitive
to age as to metallicity, and roughly three times more sensitive
to both age and metallicity than $M-T_1$ (Cellone \& Forte,
1996). One thus expects younger objects to have bluer $C-T_1$. The
M and $T_1$ bands are equivalent to the Johnson-Cousins V and R
bands, respectively. The M-band is centered on ~500 nm, and
includes the OIII 5007, OIII 4959, and H$\beta$ lines. The $T_1$
band is centered on the H$\alpha$ line, and on the [NII] 6548 and
6584 lines which are strong and in emission in planetary nebulae (PNe).
The increase of $C-T_1$ may also be caused by the presence of
PNe (and thus of Balmer emission lines).
PNe are rare in GCs, as they are created during the final
stages of the life of stars whose birth masses were between 1 and
8 M$\sun$; however, one can statistically expect a larger number of
massive stars and thus of PNe in more massive GCs.
There are only 4 known PNe in
Galactic GCs (Jacoby et al. 1997) and there is one confirmed
PN in the GCs of NGC 5128 (Minniti \& Rejkuba 2002)
and a few candidates (Rejkuba, Minniti \& Walsh 2003). On the
other hand, younger GCs have more PNe because of the higher range of
progenitor masses (see also Larsen \& Richtler 2006).
(The authors thank the referee for the above discussion on
planetary nebulae.)

Summarizing, metallicity  increases the $C-T_1$ and $M-T_1$ colors
in such a way that objects move along the reddening line on the
color-color diagram, which is parallel to the SSP tracks of
varying metallicity. Systematically redder $C-T_1$ at a given
metallicity indicate older ages, redder horizontal branches, or
the influence of emission-line objects on the integrated colors.
This is why cores of disrupted dwarf galaxies, containing multiple
stellar populations, may have bluer $(M-T_1)$ and redder $C-T_1$
colors.

The GCs of G1 and G3 seem to be located roughly in the same region
of the color-color diagram, but the scatter in G3 is large,
presumably because the photometric uncertainties on these fainter
objects are larger, and precludes any definite interpretation.

The GCs of G2 and a subset of LSB dwarf galaxies stand out in
Fig.8 : most of them lie on a track of younger age and/or of
higher metallicity than the GCs of G1 and the Galactic GCs.
Cellone \& Forte (1996) interpret the "deviating branch" of LSB
dwarf galaxies as caused by a mixture of stellar populations,
including younger components. We adopt this interpretation for G2
and argue that the GCs of this group, which are the most massive
GCs, have several generations of stars.

This property is shared by a growing number of massive GCs in our
own Galaxy (Piotto, 2009 and references therein). However, these
galactic GCs are metal-poor, while G2 is composed mostly of
metal-rich GCs.  Furthermore, galactic GCs appear to be
intermediate between G2 and G1 in Fig 8; their average mass is 5.2
$\pm$ 0.6 in log(M/M$_\odot$), using the mass estimates of
McLaughlin \& Van Der Marel (2005), thus closer to G1 than to G2.
In other words, the GCs in G2 bear little resemblance to the
galactic GCs, and their large mass presumably allowed for multiple
generations of stars more like what occurs in galaxies, thanks to
their large potential well which retained the metals lost to the
stars.

The kinematic properties of the different groups may
provide clues to their origin. G1 rotates in the same way as the
majority of GCs and PNe of the galaxy (Woodley et
al. 2007 and references therein), but has a lower mean velocity
dispersion (102.1 km/s instead of $\simeq$ 110 km/s). G2 shows
no significant mean rotation and a higher velocity dispersion than
G1, comparable to that found by Woodley et al. (2007) for the GCs
in the outer regions of the galaxy. The mean radial velocities of
G1 and G2 are significantly different: that of G1 is close to the
mean recession velocity of the galaxy, whereas that of G2 is much
larger.  This can be seen in Fig.9, which shows the radial
velocity of the GCs against their position angle (measured from
north eastward). This figure also shows that the rotation of G1 is
marginal. Only one GC of G3 has a measured radial velocity. The
gaps in position angle are due to the fact that the structural
parameters were measured on several images which do not cover the
galaxy uniformly.

\subsection{Distinct origins of the three groups of globular clusters}

We now proceed to interpret the distinctive properties of the
three groups in view of explaining their possibly different
origins; this must be done in the framework of the evolutionary
history of the galaxy itself. We further have to assume that the
structural characteristics of the GCs, on which this whole study
is based, are indeed appropriate for discriminating between
different histories of formation of GCs. While $r_h$ remains
constant, all the other structural parameters change during the
evolution. Furthermore, numerical simulations have shown that mass
segregation and loss of low-mass stars have an effect on both the
photometric and structural properties of GCs (e.g. Lamers et al.
2006). The distinct properties of the groups described in the
preceding section do confort our working hypothesis, especially
the fact that G2 is deviant in the color-color plot shown on Fig.
8.

NGC 5128 is an elliptical galaxy with a rotating dust lane
(Graham, 1979) and a system of shells (Malin et al. 1983), both of
which are characteristics of a past merger event, but not
necessarily a unique one. Estimates for the age of the merger(s)
range from 200 Myr to several Gyr (see Israel, 1998 for a review).
The gaseous component associated with the dust lane is in rapid
rotation about the major axis of the galaxy, in the same sense as
G1, with a maximum rotation of 200 km/s for the ionized gas (Bland
et al. 1987) and 265 km/s for the HI gas (van Gorkom et al. 1990).
The gaseous disk is warped, and the gas is still unstable in the
outer regions, two indications that the merger is recent.  The
stellar component of the galaxy rotates around the minor axis,
with a maximum velocity of about 40 km/s (Wilkinson et al. 1986).

Considering first G2, it has no net rotation, is composed of old,
massive, metal-rich and dynamically evolved GCs, and, because of
its high mean radial velocity, might be associated with a
high-velocity component of molecular gas (Israel 1998 and
references therein).  We propose that this group formed during the
very first merger that gave rise to the elliptical galaxy. It is
interesting to note that the metallicity distribution and age of
G2 are very close to those of the outer halo stars of NGC 5128,
for which Rejkuba et al. (2005) derived a mean metallicity of
-0.64 and a mean age of 8 Gyr. It is thus tempting to assume that
the GCs of G2 and the outer halo stars have the same origin.

The GCs of G2 are on average 9.4 Gyr old (see Table 4); however,
considering that these are mostly massive GCs, this age might be
influenced by the unwanted presence in the spectra of
horizontal-branch stars, and in fact much older. Consequently, the
merger may be even older than 9.4 Gyr, and at any rate much older
than the one(s) that gave rise to the shells and dust lane.
Mergers do produce a large number of super star clusters; for
example, a thousand such clusters have been detected in NGC
4038/4039 (Mengel et al. 2005), a minority of which could later
become massive GCs (Whitmore et al. 2007).  Super star clusters in
that and other recent merger remnants, NGC 7252 (Schweizer \&
Seitzer 1998), NGC 1275 (Zepf et al. 1995), are generally of solar
metallicity. The metallicity of G2 is lower than solar, because
the merger occurred at least 9.4 Gyr ago. Since G2 is composed of
the most massive GCs of the sample, their high mass may be the
cause of the multiple star formation episodes, more massive GCs
being able to retain their metals for future generations of stars.
A hint of a mass-metallicity relation can indeed be seen in
Fig.10, if one ignores the low-mass GCs (those of G3), in the
sense that there are no massive and metal-poor GCs in NGC 5128.

G3, on the contrary, is composed of low-mass, dynamically young
GCs, and populates preferentially the external regions of the
galaxy. This suggests that the GCs of this group were formed in
satellite dwarf galaxies which were later progressively accreted
into the halo of the galaxy and subsequently disrupted. It is
likely that most accretion events occurred recently, since their
probability increases with the mass of the attracting galaxy. The
metallicity range of the metal-poor component of G3 points to
progenitor galaxies of luminosities in the range $10^8 - 10^9
L_{\odot}$, assuming that they have the same metallicity and a
standard luminosity-metallicity relation (e.g. Lamareille et al
2009). The accretion and disruption of such low-mass galaxies
could also explain a tidal stream of young stars discovered in the
halo of NGC 5128 (Peng et al. 2002). Such accretion events may
substantially increase the number of GCs in NGC 5128, because the
number of GCs per unit galaxy mass ($S_N$) can be very high in the
low-mass galaxies of the Centaurus group (e.g. $S_N$ = 100 for
KK221, Puzia \& Sharina, 2008). The metal-rich component of G3
might have originated in the accretion of slightly larger
galaxies, like the LMC (see below), or, alternatively, be the
low-mass end of G1 and/or G2.

The GCs of G1 have properties which are intermediate between those
of G2 and G3, in mass and dynamical evolution. They are also
marginally younger than those of G2. Furthermore, these GCs rotate
in the same sense as the gaseous component, but much more slowly
so. They might thus have originated in a galaxy that merged with
NGC 5128, giving rise to the dust lane and maybe also the shells,
and that is still in the process of settling dynamically in the
merger remnant. Dynamical friction might be the cause of the
slower velocity of rotation of the GCs of G1 compared to the
gaseous component.

The peak at [Fe/H] = - 1 in all three groups may be an indication
of a significant contribution of $10^{9}$ to $10^{10} M_{\odot}$
haloes during the hierarchical galaxy formation, assuming that
these haloes produced GCs with roughly the same metallicity and
that they follow the mass-metallicity relation at z $\simeq$ 1
(Lamareille et al. 2009). For comparison, we mention that the
metallicity distribution of M87, another giant elliptical galaxy
with an active nucleus, also peaks around - 1 (Cohen et al.1998),
but, unlike NGC 5128, does not have a very low metallicity tail.
There are in fact other analogies between the metallicity
distributions of the three groups and those of GCs in nearby
galaxies : for example between the GCs of M 31 (Barmby et al.
2000) and G1, or the GCs of the LMC (Beasley et al. 2002) and the
metal-rich subgroup of G3. However, these comparisons can only
give order of magnitude estimates, since the LMC may have been
metal-enriched by interactions with our Galaxy, and M 31 might
have a very different evolutionary history from the progenitor
galaxy of G1, which merged at least 200 Myrs ago with NGC 5128.

In summary, the above scenario rests mainly on the assumption of
accretion events which shaped the evolution of NGC 5128 and its
GCs. The latter have several possible origins : the GCs of G2 were
produced in a major merger, while the GCs of the two other groups
were pre-existing in smaller galaxies that were subsequently
accreted and disrupted. This favors the categories ii) and v)
listed in the introduction for the formation of the galaxy itself,
namely a major merger and several accretions and in-situ merging.
The proposed scenario remains highly speculative, in the absence
of spectroscopically determined ages and metallicities for most
GCs in the galaxy.

\section{Acknowledgements}

TC and AKC are  thankful to IUCAA for assistance through
Associateship Programme. MES acknowledges of a partial support of
a Russian Foundation Basic Research grant 08-02-00627. The authors
are extremely grateful to Ethan T. Vishniac, Editor in Chief and
Richard de Grijs (Scientific Editor) for their active cooperation
and support. The authors are also very thankful to the referee for
valuable suggestions and technical guidance in improving the
quality of the work.

\appendix

\section{PCA based on Multivariate MM-estimator with Fast and Robust Bootstrap}

This method was developed by Salibi\'an-Barrera
et al.(2006). This Principal Component Analysis (PCA) is new in the
sense that it is based on multivariate MM-estimator of shape
instead of sample covariances. MM-estimator gives a robust
estimate having the properties of maximum likelihood estimator. A
robust estimate is not affected by outliers or small deviations
from the model assumptions. The definitions of MM-estimators of
Multivariate location and shape are given in the above mentioned
paper. The-MM estimators were computed by applying the fast
bootstrap procedure of Salibi\'an-Barrera \& Zamar (2002). If
there are $p$ parameters in the data set and we denote the
estimated variances of $p$ principal components by
$\widehat{\lambda_{1}} > \widehat{\lambda_{2}} > ... >
\widehat{\lambda_{p}}$, then to find the optimum number of
principal components the following proportion has been used
(Salibi\'an-Barrera et al. 2006) :
\begin{equation}
\begin{array}{ccc}
\widehat{p}_{k} =
\frac{\sum\limits_{j=1}^{k}\widehat{\lambda}_{j}}
{\sum\limits_{j=1}^{p}\widehat{\lambda}_{j}}, & for & k = 1, ...,
p - 1.
\end{array}
\end{equation}
One should consider that value of k as optimum for which the value
of 100$\widehat{p}_{k}$ exceeds some cut off value. In our case,
this cut off value has been chosen as 90\%. One can also test the
hypothesis whether the actual proportion ${p}_{k}$ exceeds the cut
off value 0.90\% on the basis of an one sided confidence interval
for ${p}_{k}$ based on the fast bootstrap.

\section {K-Means and mixture model methods of cluster analysis}

The target of the mixture model method developed by Qui \& Tamhane (2007) is
to divide $n$ observations into $K (<n)$ clusters so that the
within cluster variations are very small and any type of
dissimilarity must occur between clusters. In this work Qui \&
Tamhane (2007) have assumed K to be known. In our analysis, we have
computed the value of K by K-means clustering (MacQueen 1967)
and a statistical technique developed by Sugar \& James (2003).
By using this algorithm we have first determined the structures of
subpopulations (clusters) for varying number of clusters taking
K = 1, 2, 3, 4 etc. For each such cluster formation we have computed
the values of a distance measure $d_{K} = (1/p) min_{x} E[( x_{K}
- c_{K} )^{'}(x_{K} - c_{K})]$ which is defined as the distance of
the $x_{K}$ vector (values of the parameters) from the center
$c_{K}$ (which is estimated as mean value), p is the order of the
$x_{K}$ vector. Then the algorithm for determining the optimum
number of clusters is as follows (Sugar \& James 2003). Let us
denote by $d_{K}^{\bf'}$ the estimate of $d_{K}$ at the $K^{th}$
point. Then $d_{K}^{\bf'}$ is the minimum achievable distortion
associated with fitting K centers to the data. A natural way of
choosing the number of clusters is to plot $d_{K}^{\bf'}$ versus K
and look for the resulting distortion curve. This curve is always
monotonic decreasing. Initially one would expect much smaller
drops i.e. a levelling off for K greater than the true number of
clusters because past this point adding more centers simply
partitions within groups rather than between groups. According to
Sugar \& James (2003) for a large number of items the distortion
curve when transformed to an appropriate negative power ($p/2$ in
our case), will exhibit a sharp "jump" (if we plot K versus
transformed $d_{K}^{\bf'}$). Then we have calculated the jumps in
the transformed distortion as
 $J_{K} = (d_{K}^{\bf'-(p/2)}  - d_{K-1}^{\bf'-(p/2)}$).
The optimum number of clusters is the value of  K at which the
distortion curve levels off as well as its value associated with
the largest jump.
  \newline
  In this Mixture Model method the data has been considered as
  a mixture of k Multivariate distributions with unknown mixing
  proportions. The author have used EM algorithm to estimate the
  parameters of the Multivariate distributions considered in the
  model as well as the unknown mixing proportions. The method of
  maximization of the likelihood function (which is a part of the
  EM Algorithm) has been derived by McLachlan \& Krishnan (1997).

\section{The Levenberg-Marquardt algorithm}

In the present paper we have determined the rotation amplitudes
($\Omega$R) and the position angles of the axes of rotation (East
of North) of different groups of GCs ($\psi_{0}$) obtained from
the classification of GCs of NGC 5128. For this we have used the
relation
\begin{equation}
v_{r}(\psi) = v_{sys} + \Omega Rsin(\psi - \psi_{0})
\end{equation}
(C$\widehat{o}$t\'e et al. 2001; Richtler et al. 2004; Woodley et al.
2007). In the above equation, $v_{r}$ is the observed radial
velocity of the GCs in the system, $v_{sys}$ is the galaxy's
systematic velocity, R is the projected radial distance of each GC
from the center of the system assuming a distance of 3.8 Mpc to
NGC 5128, and $\psi$ is the projected azimuthal angle of the GC
measured in degrees east of north. The systematic velocity of NGC
5128 is held constant at $v_{sys}$ = 541 km$s^{-1}$ (Hui et al.
1995) for all kinematic calculations. The rotation axes of the
different groups of GCs, $\psi_{0}$ and the product $\Omega$R, the
rotation amplitudes are the values obtained from the numerical
solution. We have used the Levenberg-Marquardt non-linear fitting
Method (Levenberg 1944, Marquardt 1963) to solve the above
equation.

The projected velocity dispersion is calculated using
\begin{equation}
\sigma_{v}^{2} = \sum_{i=1}^{N}\frac{(v_{fi} - v_{sys})^{2}}{N}
\end{equation}
(Woodley et al. 2007), where N is the number of clusters in each
group of GCs, found as a result of the CA, $v_{fi}$ is the GC's
radial velocity after subtraction of rotational component
determined with equation (C1) and $\sigma_{v}$ is the projected
velocity dispersion.

\clearpage

\begin{figure}
\epsscale{.80}\includegraphics[angle=270, width=20cm]{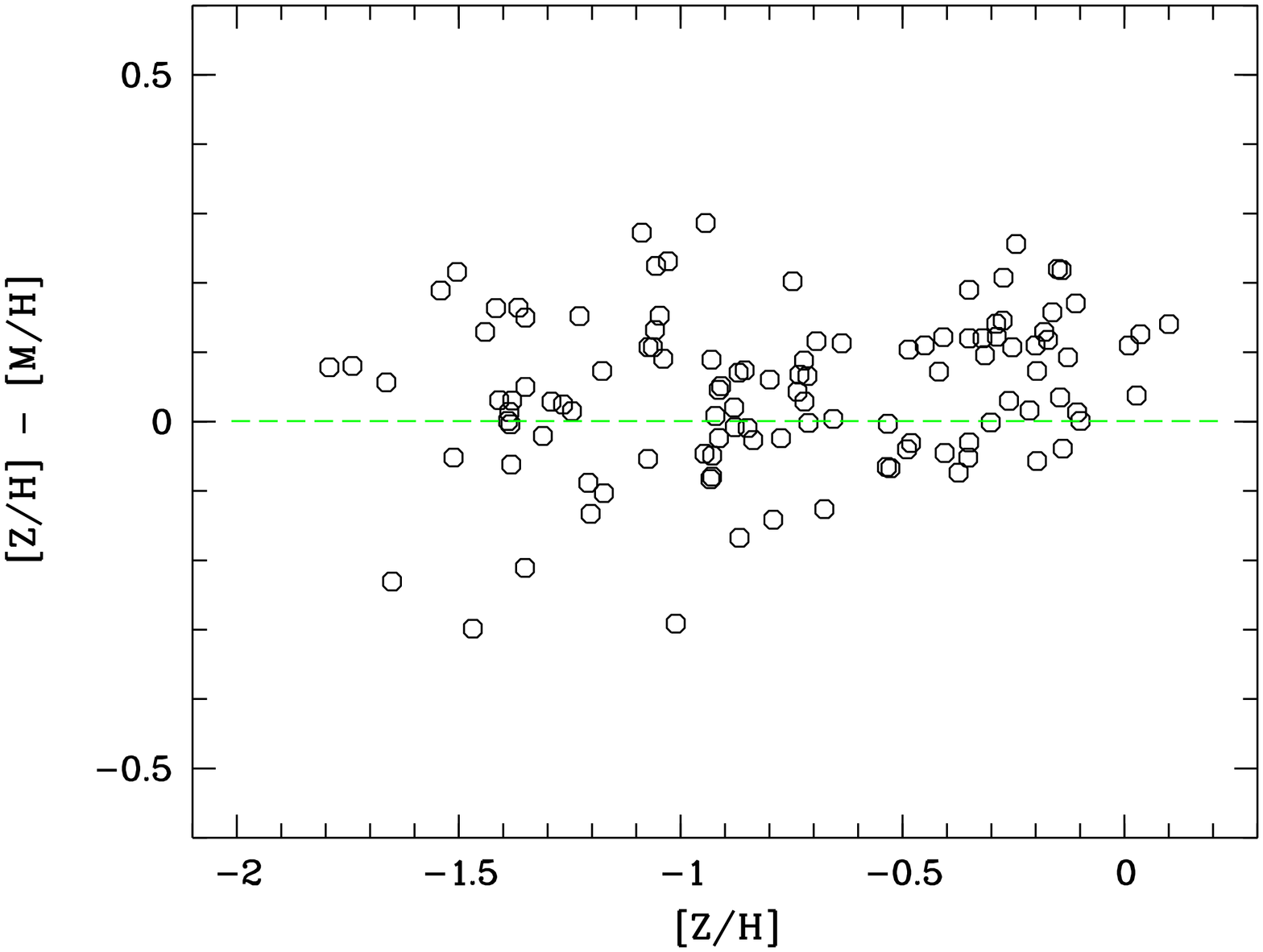} \caption{Comparison of the
metallicities ([Z/H]) derived by the present method with those of
([M/H]) derived by Beasley et al. (2008)}
\end{figure}

\clearpage

\begin{figure}
\epsscale{.80} \plotone{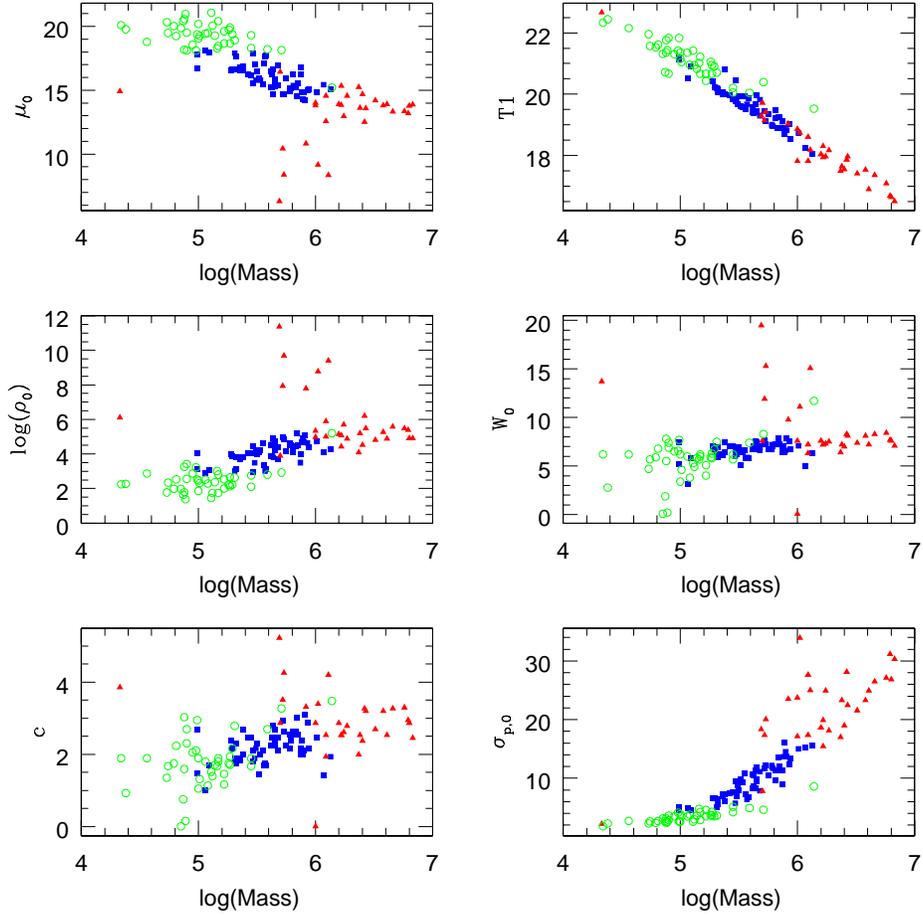} \caption{Binary diagrams of
log(mass) vs $\mu_{0}$, $T_{1}$, log($\rho_{0}$), $W_{0}$, c,
$\sigma_{p,0}$ for three groups. Blue solid squares are for G1,
red solid triangles are for G2 and green empty circles are for
G3.}
\end{figure}

\clearpage

\begin{figure}
\epsscale{.80} \plotone{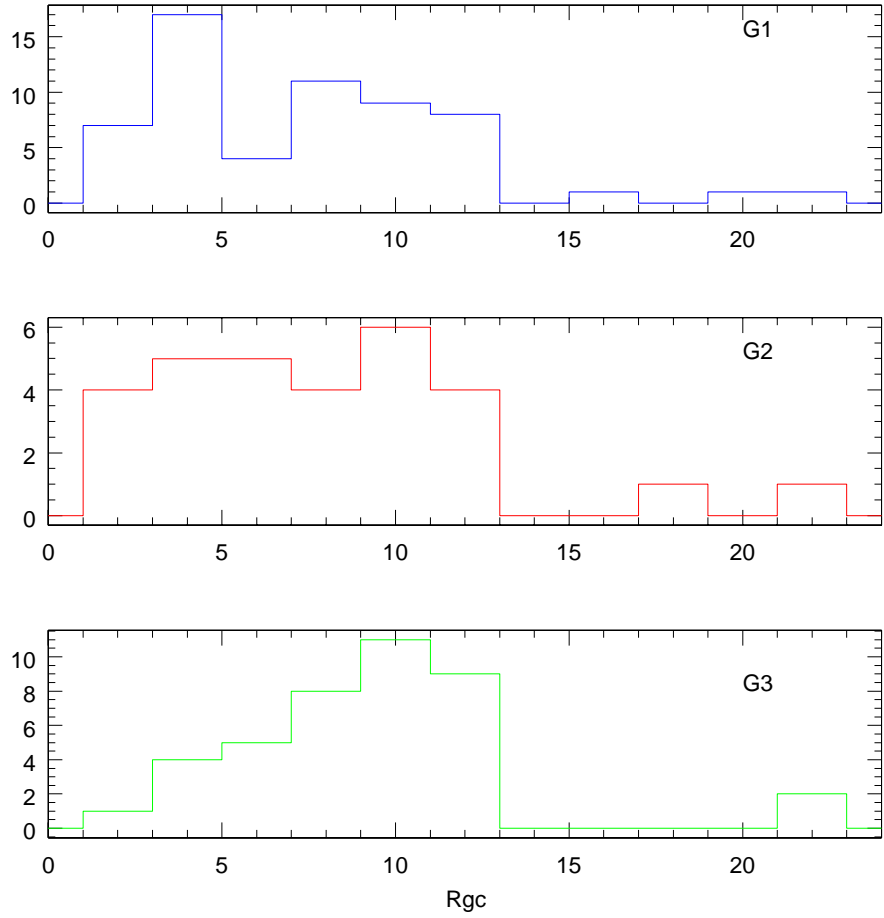} \caption{Histograms of $R_{gc}$
(galactocentric distance) for three groups. Colors are the same as in
Fig2.}
\end{figure}

\clearpage

\begin{figure}
\epsscale{.80} \plotone{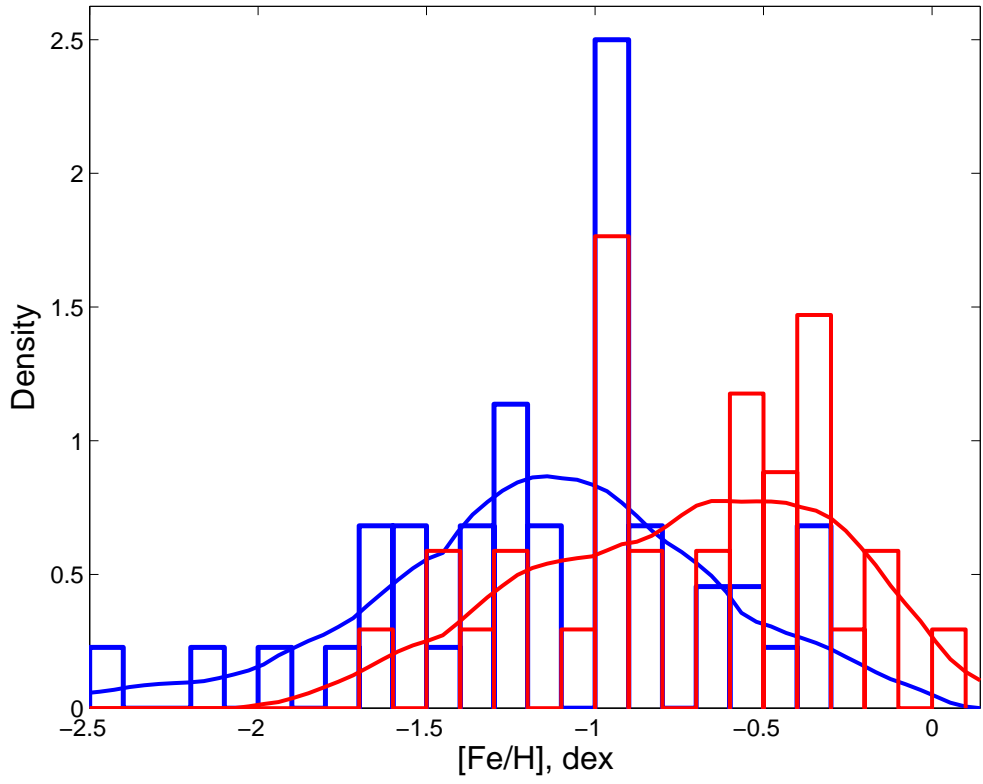} \caption{Probability density
functions of [Fe/H] for the groups G1 and G2. Colors are the same as
in Fig2.}
\end{figure}

\clearpage

\begin{figure}
\epsscale{.80} \plotone{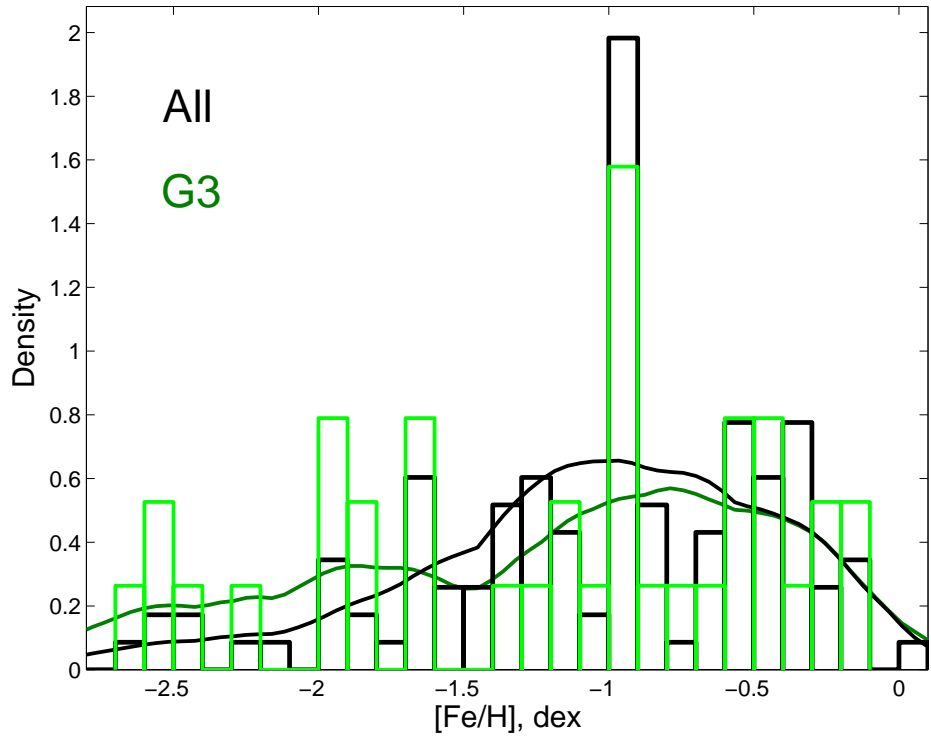} \caption{Probability density
functions of [Fe/H] for the group G3 and the whole sample. Green
is for G3 and black is for the entire sample.}
\end{figure}

\clearpage

\begin{figure}
\epsscale{.80} \plotone{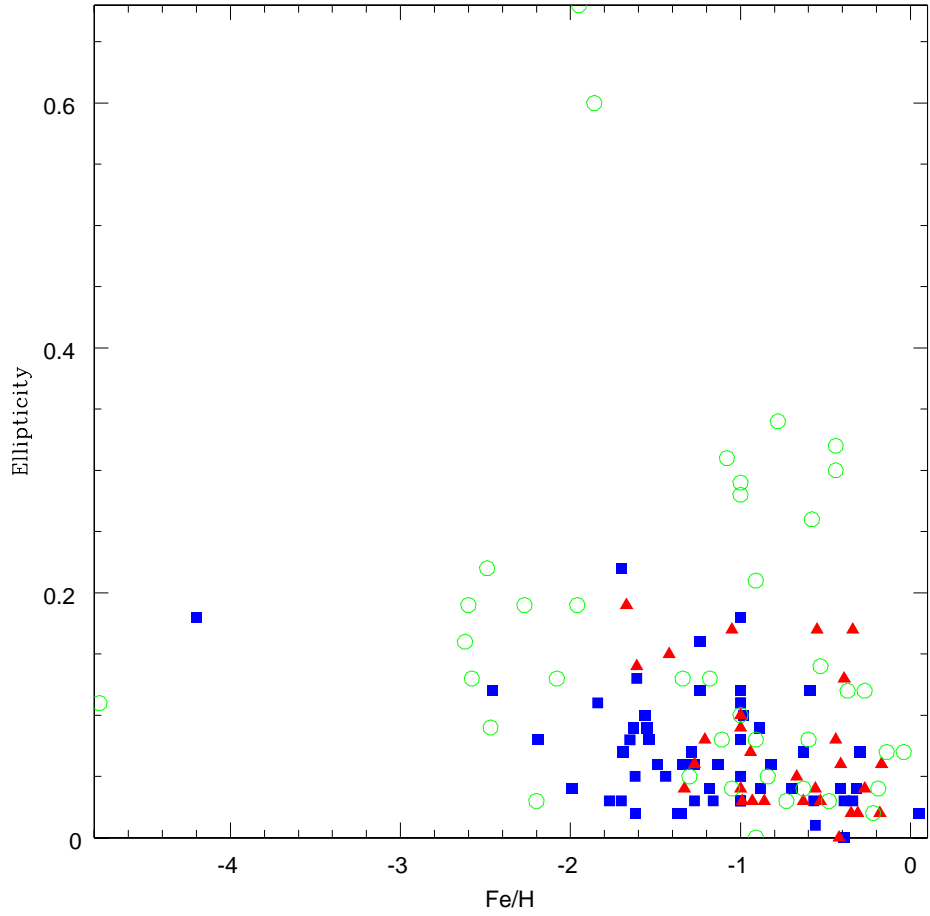} \caption{Binary plot of
ellipticity vs [Fe/H] for the three groups. Colors and symbols are
the same as in Fig.2.}
\end{figure}

\clearpage

\begin{figure}
\epsscale{.80} \plotone{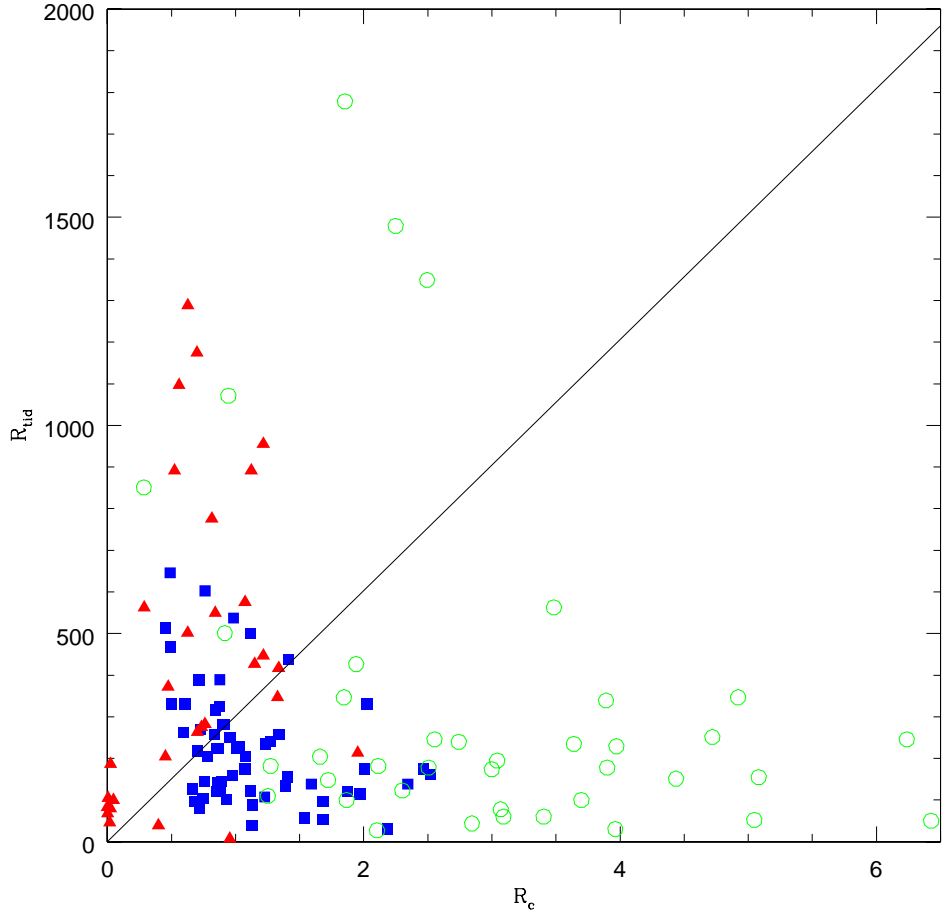} \caption{Color-color [Binary plot
of tidal radii ($R_{tid}$) vs core radii ($R_{c}$) for the three
groups. Colors and symbols are the same as in Fig2.}
\end{figure}

\clearpage

\begin{figure}
\epsscale{.65} \includegraphics[angle=270, width=20cm]{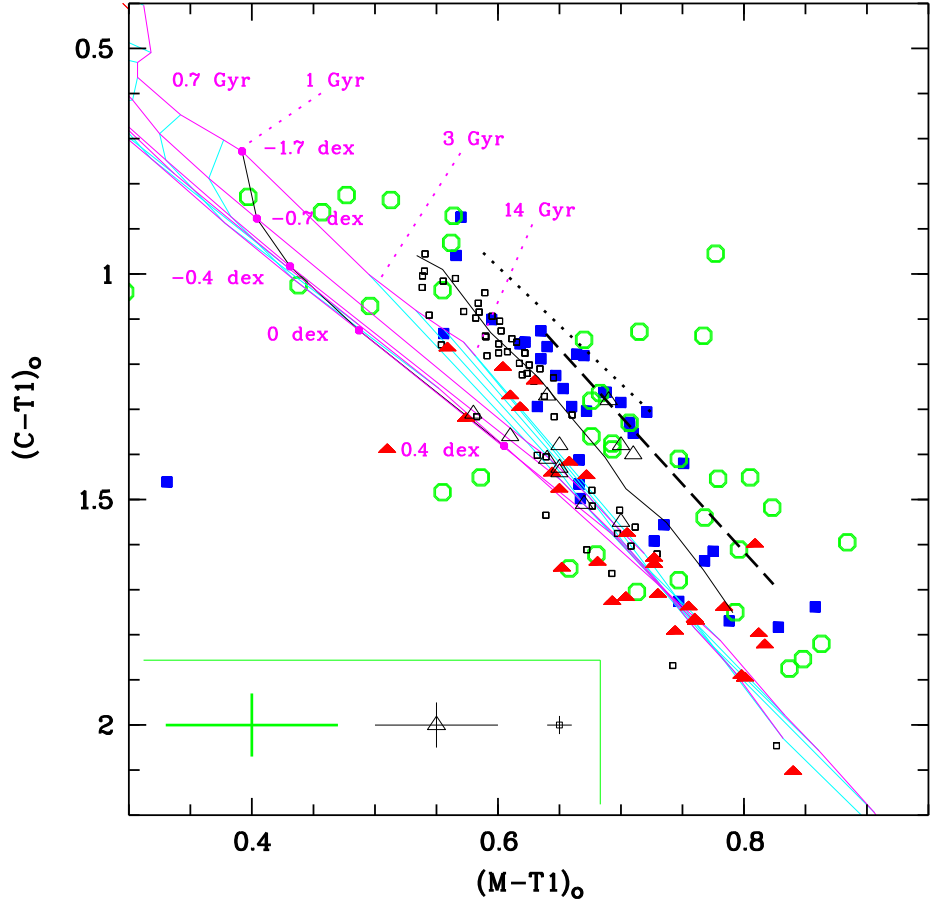} \caption{(M - $T_{1}$) vs (C -
$T_{1}$)] plot in Washington photometry for the three groups of
GCs of NGC 5128 (blue squares for G1, red triangles for G2 and
green open circles for G3), LSB dwarf galaxies (open triangles),
Galactic globular clusters (small open squares), SSP track of
varying metallicity for an age of 15Gyr from Cellone \& Forte
(1996) (solid line), tracks for E galaxies (short-dashed line) and
Sa galaxies (dotted line) from Buzzoni (2005). GALEV model grids
(ages: from 140 Myr to 14 Gyr; metallicities: z=0.0004, 0.004,
0.008, 0.02, and 0.04) are over plotted (Anders \& Fritze - v.
Alvensleben, 2003). Seven GCs fall outside the plotted range,
which has been reduced for clarity. The error bars are, from left to right,
$\sigma (C-T_1) = \pm 0.017 and \sigma (M-T_1) = \pm 0.009$
for Harris \& Canterna (1977),
$\pm 0.07$ in both colors for Harris et al.(2004), and $\pm 0.05$
in both colors for Cellone et al. (1994).}
\end{figure}

\clearpage

\begin{figure}
\epsscale{.80} \plotone{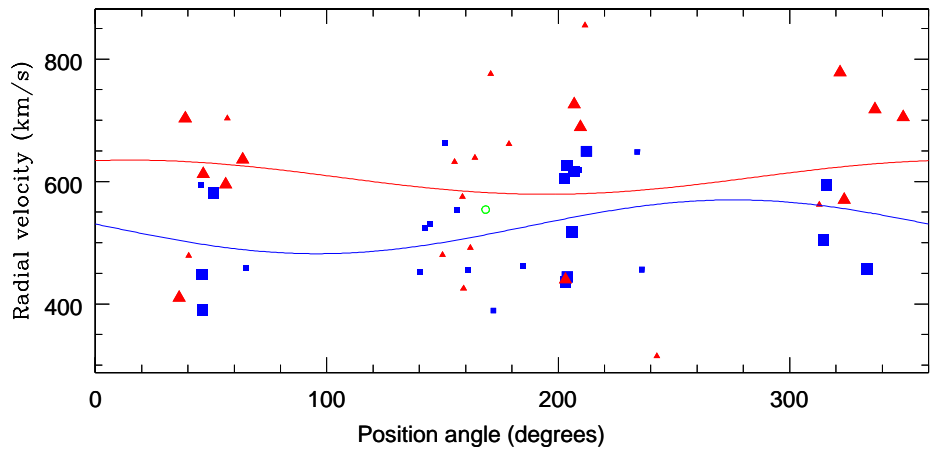} \caption{Radial velocity plotted
against position angle (measured from North eastward) for the
three groups (blue squares for G1, red triangles for G2, green
open squares for G3). larger symbols denote radii larger than 8.0
arcmin. The mean rotation curves for the two first groups are
shown as solid blue and red lines. The GCs of G2 have on average
larger radial velocities than those of G1, their mean rotation
curve is thus above that of G1. }
\end{figure}

\clearpage

\begin{figure}
\epsscale{.80} \plotone{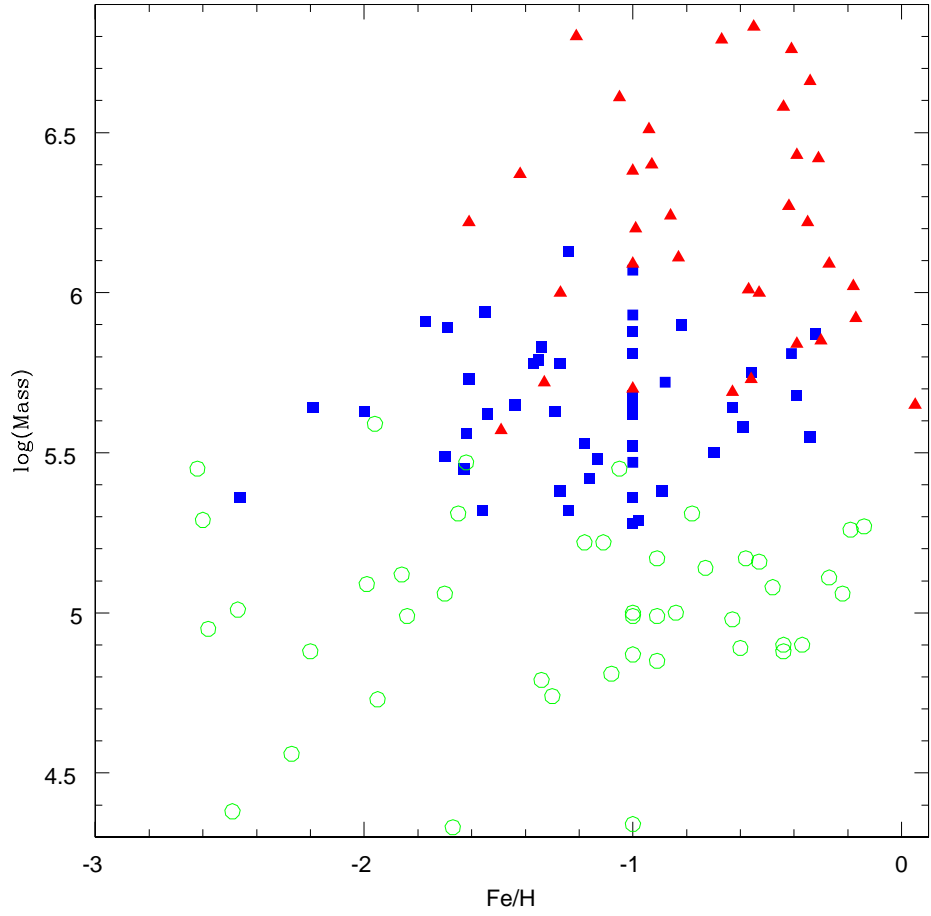} \caption{Mass-metallicity relation
for the three groups (blue squares for G1, red
triangles for G2, green open squares for G3).}
\end{figure}

\clearpage
\begin{deluxetable}{ccccccccccccc}
\tabletypesize{\scriptsize} \rotate \tablecaption{List of
parameters of the globular clusters of NGC 5128 used in PCA and
CA} \tablewidth{0pt} \tablehead{
\colhead{Name} & \colhead{Group} & \colhead{c} & \colhead{$<\mu_{v}>_{h}$} & \colhead{$\mu_{0}$} & \colhead{$W_{0}$} & \colhead{log$R_{c}$} & \colhead{log$R_{tid}$} & \colhead{log$r_{h}$} & \colhead{log$M_{tot}$} & \colhead{$\rho_{0}$} & \colhead{$\sigma_{p,0}$} & \colhead{$t_{rh}$} \\
\colhead{} & \colhead{} & \colhead{} & \colhead{(mag $arcsec^{-2}$)} & \colhead{(mag $arcsec^{-2}$)} & \colhead{} & \colhead{(pc)} & \colhead{(pc)} & \colhead{(pc)} & \colhead{($M_{\bigodot}$)} & \colhead{($M_{\bigodot}pc^{-3}$)} & \colhead{(km/sec)} & \colhead{($10^{9}$ years)}
}
\startdata
AAT111563 & 1 & 1.89 & 18.15 & 16.76 & 6.2 & 0.09 & 2.03 & 0.49 & 5.36 & 5129 & 6.90 & 0.91 \\
AAT113992 & 1 & 2.46 & 19.10 & 16.97 & 7.1 & 0.15 & 2.64 & 0.75 & 5.68 & 5012 & 7.85 & 3.16 \\
AAT115339 & 1 & 2.00 & 17.33 & 15.78 & 6.4 & -0.03 & 2.01 & 0.40 & 5.53 & 16596 & 9.38 & 0.78 \\
AAT119508 & 1 & 1.71 & 19.03 & 17.70 & 5.8 & 0.37 & 2.14 & 0.71 & 5.58 & 1380 & 6.84 & 2.51 \\
AAT120336 & 1 & 2.24 & 18.50 & 16.66 & 6.8 & 0.13 & 2.41 & 0.64 & 5.64 & 6166 & 8.28 & 2.09 \\
AAT120976 & 1 & 2.00 & 18.08 & 16.53 & 6.4 & 0.05 & 2.09 & 0.48 & 5.38 & 6761 & 7.21 & 0.91 \\
C104 & 1 & 2.24 & 17.00 & 15.21 & 6.8 & -0.12 & 2.16 & 0.39 & 5.62 & 33113 & 10.74 & 0.85 \\
C115 & 1 & 1.71 & 18.35 & 17.09 & 5.8 & 0.29 & 2.06 & 0.64 & 5.56 & 2239 & 7.29 & 1.86 \\
C123 & 1 & 2.17 & 18.38 & 16.64 & 6.7 & -0.01 & 2.20 & 0.48 & 5.29 & 7586 & 6.62 & 0.83 \\
C130 & 1 & 2.46 & 18.23 & 16.17 & 7.1 & -0.04 & 2.45 & 0.56 & 5.45 & 11220 & 7.55 & 1.26 \\
C133 & 1 & 2.11 & 17.08 & 15.40 & 6.6 & -0.06 & 2.10 & 0.41 & 5.68 & 25704 & 11.09 & 0.95 \\
C138 & 1 & 2.46 & 18.38 & 16.27 & 7.1 & -0.05 & 2.45 & 0.55 & 5.42 & 10965 & 7.34 & 1.20 \\
C140 & 1 & 1.89 & 18.48 & 17.02 & 6.2 & 0.20 & 2.14 & 0.60 & 5.48 & 3090 & 6.97 & 1.51 \\
C146 & 1 & 2.31 & 18.38 & 16.45 & 6.9 & 0.01 & 2.36 & 0.55 & 5.50 & 9333 & 7.85 & 1.32 \\
C147 & 1 & 1.75 & 18.98 & 17.65 & 5.9 & 0.27 & 2.08 & 0.63 & 5.32 & 1445 & 5.58 & 1.41 \\
C149 & 1 & 2.17 & 17.48 & 15.78 & 6.7 & -0.06 & 2.15 & 0.42 & 5.49 & 16982 & 8.83 & 0.83 \\
C150 & 1 & 2.11 & 17.25 & 15.57 & 6.6 & -0.13 & 2.02 & 0.34 & 5.47 & 26303 & 9.46 & 0.62 \\
C154 & 1 & 2.24 & 18.25 & 16.45 & 6.8 & 0.11 & 2.38 & 0.62 & 5.62 & 6918 & 8.26 & 1.86 \\
C157 & 1 & 2.54 & 17.58 & 15.36 & 7.2 & -0.06 & 2.51 & 0.58 & 5.81 & 26915 & 11.30 & 1.90 \\
C159 & 1 & 2.11 & 17.23 & 15.51 & 6.6 & -0.17 & 1.98 & 0.30 & 5.55 & 40738 & 10.79 & 0.59 \\
C160 & 1 & 1.84 & 18.00 & 16.58 & 6.1 & 0.05 & 1.94 & 0.44 & 5.36 & 6761 & 7.33 & 0.76 \\
C164 & 1 & 1.46 & 17.05 & 15.95 & 5.1 & 0.05 & 1.59 & 0.33 & 5.52 & 12023 & 9.84 & 0.62 \\
C167 & 1 & 2.38 & 18.53 & 16.54 & 7.0 & -0.06 & 2.35 & 0.50 & 5.28 & 9333 & 6.50 & 0.89 \\
F2GC14 & 1 & 2.70 & 19.33 & 16.84 & 7.4 & -0.01 & 2.73 & 0.70 & 5.38 & 6457 & 6.15 & 1.95 \\
F2GC31 & 1 & 2.31 & 18.53 & 16.64 & 6.9 & 0.004 & 2.35 & 0.54 & 5.32 & 6607 & 6.41 & 1.07 \\
PFF034 & 1 & 2.24 & 18.05 & 16.26 & 6.8 & 0.09 & 2.37 & 0.60 & 5.63 & 7943 & 8.55 & 1.78 \\
PFF041 & 1 & 2.54 & 17.13 & 14.94 & 7.2 & -0.14 & 2.43 & 0.49 & 5.78 & 44668 & 11.97 & 1.38 \\
PFF052 & 1 & 2.70 & 17.80 & 15.34 & 7.4 & -0.15 & 2.59 & 0.56 & 5.65 & 30903 & 9.86 & 1.58 \\
PFF059 & 1 & 2.62 & 18.63 & 16.24 & 7.3 & 0.05 & 2.70 & 0.72 & 5.81 & 12303 & 9.68 & 3.16 \\
PFF063 & 1 & 2.79 & 17.28 & 14.70 & 7.5 & -0.30 & 2.52 & 0.46 & 5.64 & 83176 & 11.30 & 1.07 \\
PFF066 & 1 & 2.38 & 17.83 & 15.83 & 7.0 & -0.02 & 2.40 & 0.55 & 5.65 & 16218 & 9.48 & 1.51 \\
PFF083 & 1 & 2.95 & 17.70 & 14.69 & 7.7 & -0.31 & 2.67 & 0.54 & 5.72 & 93325 & 11.72 & 1.58 \\
R203 & 1 & 2.46 & 17.25 & 15.20 & 7.1 & -0.15 & 2.34 & 0.45 & 5.63 & 36308 & 10.52 & 1.02 \\
C113 & 1 & 2.62 & 16.98 & 14.67 & 7.3 & -0.23 & 2.42 & 0.44 & 5.73 & 69183 & 12.16 & 1.10 \\
C132 & 1 & 2.62 & 17.60 & 15.27 & 7.3 & -0.06 & 2.59 & 0.61 & 5.83 & 27542 & 11.32 & 2.19 \\
C137 & 1 & 2.87 & 17.83 & 15.01 & 7.6 & -0.12 & 2.78 & 0.68 & 5.93 & 43652 & 12.39 & 3.09 \\
PFF016 & 1 & 2.24 & 17.80 & 15.95 & 6.8 & 0.03 & 2.31 & 0.55 & 5.75 & 15488 & 10.45 & 1.66 \\
PFF021 & 1 & 3.10 & 17.55 & 14.20 & 7.9 & -0.31 & 2.81 & 0.65 & 5.91 & 128825 & 13.74 & 2.75 \\
PFF023 & 1 & 2.17 & 16.88 & 15.14 & 6.7 & -0.05 & 2.16 & 0.44 & 5.78 & 30903 & 12.19 & 1.15 \\
PFF031 & 1 & 2.38 & 17.15 & 15.17 & 7.0 & -0.07 & 2.35 & 0.50 & 5.79 & 30903 & 11.70 & 1.44 \\
PFF035 & 1 & 2.00 & 17.83 & 16.22 & 6.4 & 0.15 & 2.19 & 0.58 & 5.87 & 10471 & 11.25 & 2.09 \\
C043 & 1 & 1.94 & 16.58 & 15.09 & 6.3 & 0.14 & 2.13 & 0.56 & 6.13 & 20417 & 15.63 & 2.45 \\
C135 & 1 & 2.38 & 16.80 & 14.82 & 7.0 & -0.11 & 2.31 & 0.46 & 5.88 & 50119 & 13.68 & 1.38 \\
C153 & 1 & 1.43 & 16.50 & 15.42 & 5.0 & 0.23 & 1.73 & 0.49 & 6.07 & 13183 & 15.31 & 1.82 \\
G221 & 1 & 2.11 & 16.53 & 14.85 & 6.6 & -0.07 & 2.08 & 0.40 & 5.90 & 47863 & 14.52 & 1.15 \\
G293 & 1 & 2.00 & 15.78 & 14.27 & 6.4 & -0.14 & 1.90 & 0.29 & 5.89 & 81283 & 16.14 & 0.76 \\
PFF011 & 1 & 2.17 & 16.83 & 15.12 & 6.7 & 0.03 & 2.24 & 0.521 & 5.94 & 25119 & 13.30 & 1.78 \\
AAT118198 & 2 & 3.32 & 16.53 & 10.81 & 9.8 & -1.33 & 2.00 & 0.32 & 5.92 & 6.30957$\times10^{7}$ & 23.50 & 0.91 \\
C006 & 2 & 2.46 & 16.00 & 13.88 & 7.1 & 0.13 & 2.62 & 0.72 & 6.83 & 8.31764$\times10^{4}$ & 30.34 & 8.91 \\
C018 & 2 & 2.54 & 16.13 & 13.91 & 7.2 & 0.06 & 2.63 & 0.69 & 6.61 & 7.76247$\times10^{4}$ & 24.95 & 6.31 \\
C030 & 2 & 2.95 & 16.23 & 13.21 & 7.7 & -0.09 & 2.89 & 0.76 & 6.79 & 2.39883$\times10^{5}$ & 31.19 & 9.77 \\
C032 & 2 & 3.27 & 16.78 & 12.51 & 8.3 & -0.54 & 2.75 & 0.64 & 6.42 & 1.58489$\times10^{6}$ & 28.12 & 4.47 \\
C037 & 2 & 2.62 & 15.33 & 12.97 & 7.3 & -0.34 & 2.31 & 0.33 & 6.24 & 4.89779$\times10^{5}$ & 24.94 & 1.23 \\
C142 & 2 & 2.54 & 15.85 & 13.63 & 7.2 & -0.12 & 2.45 & 0.51 & 6.38 & 1.54882$\times10^{5}$ & 23.33 & 2.69 \\
C145 & 2 & 0.01 & 14.45 & 13.80 & 0.1 & -0.02 & 0.87 & 0.06 & 6.00 & 9.54993$\times10^{4}$ & 23.77 & 0.39 \\
C152 & 2 & 1.94 & 14.08 & 12.57 & 6.3 & -0.40 & 1.59 & 0.01 & 6.09 & 7.76247$\times10^{5}$ & 27.60 & 0.36 \\
F2GC69 & 2 & 5.23 & 16.60 & 6.32 & 19.5 & -3.31 & 1.92 & 0.28 & 5.69 & 2.29087$\times10^{11}$ & 18.32 & 0.63 \\
G284 & 2 & 4.26 & 16.13 & 8.37 & 15.3 & -2.42 & 1.83 & 0.20 & 5.73 & 4.78630$\times10^{9}$ & 20.09 & 0.49 \\
K131 & 2 & 3.40 & 15.28 & 9.15 & 11.1 & -1.75 & 1.66 & 0.12 & 6.02 & 6.02560$\times10^{8}$ & 33.96 & 0.50 \\
PFF079 & 2 & 3.51 & 16.70 & 10.43 & 11.9 & -1.61 & 1.90 & 0.38 & 5.72 & 8.51138$\times10^{7}$ & 17.38 & 0.89 \\
R223 & 2 & 4.20 & 15.98 & 8.35 & 15.1 & -2.19 & 2.02 & 0.39 & 6.11 & 2.51189$\times10^{9}$ & 25.00 & 1.38 \\
C117 & 2 & 2.54 & 17.80 & 15.53 & 7.2 & -0.08 & 2.50 & 0.56 & 5.85 & 3.31131$\times10^{4}$ & 12.02 & 1.90 \\
C168 & 2 & 2.87 & 19.25 & 16.43 & 7.6 & 0.05 & 2.95 & 0.85 & 5.70 & 7.94328$\times10^{3}$ & 7.82 & 4.47 \\
F1GC14 & 2 & 1.75 & 18.95 & 17.64 & 5.9 & 0.40 & 2.21 & 0.76 & 5.57 & 1.04713$\times10^{3}$ & 6.41 & 2.82 \\
PFF029 & 2 & 2.17 & 18.40 & 16.80 & 6.7 & 0.31 & 2.52 & 0.80 & 5.87 & 3.16228$\times10^{3}$ & 8.95 & 4.36 \\
C161 & 2 & 3.03 & 17.78 & 14.54 & 7.8 & -0.34 & 2.71 & 0.56 & 5.84 & 1.47911$\times10^{5}$ & 13.61 & 1.95 \\
C003 & 2 & 3.30 & 17.80 & 13.37 & 8.4 & -0.20 & 3.11 & 1.03 & 6.76 & 3.09030$\times10^{5}$ & 27.10 & 23.44 \\
C004 & 2 & 2.00 & 16.78 & 15.25 & 6.4 & 0.29 & 2.33 & 0.72 & 6.37 & 1.23027$\times10^{4}$ & 16.98 & 5.37 \\
C007 & 2 & 2.87 & 16.58 & 13.77 & 7.6 & 0.08 & 2.98 & 0.8 & 6.80 & 7.94328$\times10^{4}$ & 26.85 & 14.79 \\
C012 & 2 & 3.27 & 3.27 & 3.27 & 3.27 & 3.27 & 3.27 & 3.27 & 3.27 & 3.27000$\times10^{0}$ & 3.27 & 3.27 \\
C014 & 2 & 2.70 & 16.68 & 14.19 & 7.40 & 0.03 & 2.76 & 0.74 & 6.51 & 6.60693$\times10^{4}$ & 21.58 & 6.76 \\
C019 & 2 & 2.38 & 16.70 & 14.70 & 7.00 & 0.12 & 2.54 & 0.69 & 6.40 & 3.38844$\times10^{4}$ & 19.01 & 5.13 \\
C025 & 2 & 3.20 & 17.45 & 13.60 & 8.10 & -0.28 & 2.95 & 0.79 & 6.43 & 3.01995$\times10^{5}$ & 22.44 & 7.58 \\
C029 & 2 & 3.20 & 17.65 & 13.80 & 8.10 & -0.16 & 3.07 & 0.92 & 6.58 & 1.81970$\times10^{5}$ & 23.28 & 13.49 \\
C036 & 2 & 2.54 & 16.03 & 13.83 & 7.20 & -0.13 & 2.44 & 0.50 & 6.22 & 1.20226$\times10^{5}$ & 19.91 & 2.19 \\
C139 & 2 & 2.87 & 16.85 & 14.00 & 7.60 & -0.32 & 2.57 & 0.48 & 6.00 & 2.13796$\times10^{5}$ & 17.18 & 1.66 \\
C151 & 2 & 2.24 & 17.43 & 15.53 & 6.80 & -0.18 & 2.10 & 0.34 & 5.65 & 5.12861$\times10^{4}$ & 11.83 & 0.76 \\
C165 & 2 & 2.79 & 17.25 & 14.57 & 7.50 & -0.08 & 2.74 & 0.68 & 6.27 & 7.58578$\times10^{4}$ & 18.11 & 4.36 \\
G170 & 2 & 2.46 & 17.00 & 14.88 & 7.10 & -0.08 & 2.41 & 0.52 & 6.01 & 5.24807$\times10^{4}$ & 15.03 & 1.95 \\
WHH09 & 2 & 2.54 & 17.63 & 15.35 & 7.20 & 0.08 & 2.65 & 0.72 & 6.22 & 2.63027$\times10^{4}$ & 15.45 & 4.79 \\
WHH16 & 2 & 2.54 & 16.83 & 14.55 & 7.20 & -0.15 & 2.42 & 0.48 & 6.09 & 1.00000$\times10^{5}$ & 17.38 & 1.82 \\
WHH22 & 2 & 2.87 & 16.73 & 13.91 & 7.60 & -0.20 & 2.70 & 0.60 & 6.20 & 1.44544$\times10^{5}$ & 18.62 & 3.02 \\
C126 & 3 & 3.86 & 21.80 & 14.93 & 13.7 & -1.59 & 2.27 & 0.71 & 4.33 & 1258925 & 2.24 & 0.74 \\
AAT117287 & 3 & 1.80 & 19.20 & 17.87 & 6.0 & 0.39 & 2.24 & 0.76 & 5.47 & 871 & 5.71 & 2.57 \\
C111 & 3 & 3.03 & 21.33 & 18.18 & 7.8 & -0.02 & 3.03 & 0.88 & 4.88 & 1820 & 3.12 & 2.24 \\
C114 & 3 & 1.52 & 22.28 & 21.08 & 5.3 & 0.80 & 2.39 & 1.08 & 5.11 & 28 & 2.58 & 5.89 \\
C118 & 3 & 2.70 & 20.65 & 18.12 & 7.4 & -0.04 & 2.70 & 0.67 & 4.90 & 2630 & 3.66 & 1.15 \\
C124 & 3 & 1.67 & 20.73 & 19.49 & 5.7 & 0.36 & 2.09 & 0.69 & 4.74 & 214 & 2.65 & 1.05 \\
C125 & 3 & 1.89 & 19.95 & 18.48 & 6.2 & 0.32 & 2.26 & 0.72 & 5.17 & 646 & 4.24 & 1.74 \\
C127 & 3 & 2.24 & 21.08 & 19.25 & 6.8 & 0.27 & 2.54 & 0.78 & 4.81 & 347 & 2.70 & 1.51 \\
C131 & 3 & 1.89 & 19.30 & 17.89 & 6.2 & 0.30 & 2.24 & 0.70 & 5.31 & 1047 & 5.09 & 1.82 \\
C136 & 3 & 1.49 & 18.95 & 17.85 & 5.2 & 0.19 & 1.75 & 0.47 & 4.99 & 1413 & 4.57 & 0.60 \\
C143 & 3 & 0.99 & 18.93 & 18.12 & 3.1 & 0.34 & 1.48 & 0.50 & 5.06 & 759 & 4.82 & 0.71 \\
C158 & 3 & 1.89 & 19.75 & 18.30 & 6.2 & 0.44 & 2.38 & 0.84 & 5.45 & 575 & 5.14 & 3.31 \\
C163 & 3 & 1.17 & 19.55 & 18.65 & 4.0 & 0.49 & 1.78 & 0.69 & 5.22 & 347 & 4.59 & 1.58 \\
C170 & 3 & 1.89 & 20.25 & 18.80 & 6.2 & 0.10 & 2.04 & 0.50 & 4.56 & 759 & 2.72 & 0.45 \\
C172 & 3 & 1.71 & 19.23 & 17.97 & 5.8 & 0.23 & 1.99 & 0.57 & 5.09 & 1175 & 4.55 & 0.93 \\
C174 & 3 & 2.11 & 19.83 & 18.12 & 6.6 & 0.10 & 2.26 & 0.57 & 4.98 & 1698 & 4.11 & 0.87 \\
C176 & 3 & 2.05 & 20.10 & 18.56 & 6.5 & 0.22 & 2.31 & 0.67 & 4.95 & 724 & 3.52 & 1.15 \\
C179 & 3 & 1.32 & 20.20 & 19.25 & 4.6 & 0.49 & 1.89 & 0.72 & 5.01 & 204 & 3.48 & 1.48 \\
F2GC18 & 3 & 2.70 & 19.20 & 16.72 & 7.4 & -0.22 & 2.52 & 0.49 & 4.99 & 11220 & 5.01 & 0.66 \\
F2GC20 & 3 & 1.67 & 19.58 & 18.28 & 5.7 & 0.27 & 2.00 & 0.60 & 5.16 & 1047 & 4.76 & 1.15 \\
F2GC28 & 3 & 1.05 & 19.90 & 19.04 & 3.4 & 0.45 & 1.64 & 0.62 & 5.00 & 288 & 3.87 & 1.05 \\
C105 & 3 & 1.59 & 21.80 & 20.57 & 5.5 & 0.59 & 2.25 & 0.90 & 4.88 & 63 & 2.45 & 2.51 \\
C112 & 3 & 0.01 & 20.53 & 19.86 & 0.1 & 0.60 & 1.48 & 0.68 & 4.85 & 95 & 3.12 & 1.12 \\
C119 & 3 & 1.94 & 20.03 & 18.67 & 6.3 & 0.41 & 2.39 & 0.82 & 5.27 & 457 & 4.28 & 2.69 \\
C120 & 3 & 0.16 & 21.68 & 20.97 & 0.2 & 0.81 & 1.70 & 0.89 & 4.89 & 24 & 2.54 & 2.40 \\
C121 & 3 & 0.93 & 20.53 & 19.78 & 2.8 & 0.32 & 1.43 & 0.47 & 4.38 & 182 & 2.28 & 0.35 \\
C122 & 3 & 1.89 & 21.53 & 20.08 & 6.2 & 0.24 & 2.17 & 0.63 & 4.34 & 178 & 1.81 & 0.59 \\
C128 & 3 & 1.71 & 21.28 & 19.91 & 5.8 & 0.60 & 2.36 & 0.94 & 5.27 & 135 & 3.65 & 4.17 \\
C129 & 3 & 1.71 & 20.60 & 19.28 & 5.8 & 0.48 & 2.24 & 0.82 & 5.14 & 234 & 3.62 & 2.34 \\
C134 & 3 & 2.79 & 21.58 & 18.92 & 7.5 & 0.35 & 3.17 & 1.11 & 5.31 & 437 & 3.66 & 7.41 \\
C141 & 3 & 1.40 & 21.08 & 20.06 & 4.9 & 0.71 & 2.19 & 0.96 & 5.12 & 56 & 3.01 & 3.72 \\
C144 & 3 & 1.35 & 21.30 & 20.33 & 4.7 & 0.57 & 2.00 & 0.81 & 4.73 & 60 & 2.28 & 1.55 \\
C148 & 3 & 1.67 & 20.50 & 19.31 & 5.7 & 0.67 & 2.40 & 1.00 & 5.45 & 129 & 4.21 & 5.89 \\
C155 & 3 & 1.89 & 21.65 & 20.19 & 6.2 & 0.59 & 2.53 & 0.99 & 5.00 & 71 & 2.58 & 3.63 \\
C162 & 3 & 2.17 & 21.05 & 19.38 & 6.7 & 0.54 & 2.75 & 1.03 & 5.29 & 162 & 3.48 & 5.62 \\
C166 & 3 & 0.76 & 21.15 & 20.41 & 1.9 & 0.70 & 1.72 & 0.82 & 4.87 & 44 & 2.68 & 1.82 \\
C171 & 3 & 1.46 & 20.73 & 19.62 & 5.1 & 0.65 & 2.18 & 0.92 & 5.22 & 102 & 3.53 & 3.55 \\
C173 & 3 & 1.75 & 21.13 & 19.74 & 5.9 & 0.56 & 2.37 & 0.92 & 5.26 & 170 & 3.72 & 3.80 \\
C175 & 3 & 2.95 & 22.38 & 19.37 & 7.7 & 0.27 & 3.25 & 1.12 & 4.99 & 331 & 2.62 & 5.75 \\
C178 & 3 & 1.75 & 21.33 & 20.02 & 5.9 & 0.48 & 2.29 & 0.84 & 4.79 & 98 & 2.37 & 1.78 \\
F1GC20 & 3 & 1.80 & 21.80 & 20.41 & 6.0 & 0.69 & 2.54 & 1.06 & 5.17 & 55 & 2.86 & 5.62 \\
F1GC21 & 3 & 1.80 & 20.85 & 19.43 & 6.0 & 0.40 & 2.25 & 0.77 & 5.06 & 324 & 3.53 & 1.86 \\
F1GC34 & 3 & 1.13 & 20.40 & 19.47 & 3.8 & 0.53 & 1.78 & 0.72 & 5.08 & 195 & 3.78 & 1.62 \\
F2GC70 & 3 & 2.70 & 20.65 & 18.21 & 7.4 & 0.40 & 3.13 & 1.11 & 5.59 & 646 & 4.93 & 9.77 \\
C116 & 3 & 2.31 & 21.45 & 19.51 & 6.9 & 0.29 & 2.63 & 0.83 & 4.90 & 355 & 2.87 & 2.00 \\
C156\tablenotemark{a} & * & 3.03 & 11.21 & 7.95 & 7.80 & -1.49 & 1.56 & -0.59 & 6.22 & 9 & 1.90 & 7.72 \\
C169\tablenotemark{a} & * & 3.27 & 22.29 & 18.13 & 8.30 & 0.31 & 3.60 & 1.49 & 5.71 & 2.93 & 0.67 & 10.61 \\
F1GC15\tablenotemark{a} & * & 3.48 & 21.52 & 15.22 & 11.70 & -0.55 & 2.93 & 1.41 & 6.14 & 5.19 & 0.94 & 10.69 \\
\enddata
\tablecomments {Values in column 1 and in columns 3 to 13 are from
 McLaughlin et al.(2008).Column 2 represents the group membership found as a
 result of CA.}
\tablenotetext{a}{These GCs have not been considered during the CA
as they are outliers.}

\end{deluxetable}

\clearpage

\begin{deluxetable}{ccccccccccc}
\tabletypesize{\scriptsize} \rotate \tablecaption{List of
parameters as well as derived parameters of the globular clusters
of NGC 5128.} \tablewidth{0pt} \tablehead{
\colhead{Name} & \colhead{Group} & \colhead{$R_{gc}$} & \colhead{$T_{1}$} & \colhead{[Fe/H]} & \colhead{$(C-T_{1})_{0}$} & \colhead{$V_{r}$} & \colhead{[Z/H]} & \colhead{Age} & \colhead{$V_{rot}$} & \colhead{$\sigma_{v}$} \\
\colhead{} & \colhead{} & \colhead{(kpc)} & \colhead{(mag)} &
\colhead{(dex)} & \colhead{(mag)} & \colhead{(km/sec)} &
\colhead{(dex)} & \colhead{(Gyr)} & \colhead{(km/sec)} &
\colhead{(km/sec)} } \startdata
AAT111563   &   1   &   11.93   &   20.05   &   -2.46   &   0.87    &   649 &   \nodata &   \nodata &43.96 for G1 & 102.08 for G1\\
AAT113992   &   1   &   4.26    &   19.82   &   -0.39   &   1.74    &   648 &   \nodata &   \nodata \\
AAT115339   &   1   &   3.82    &   19.56   &   -1.18   &   1.33    &   618 &   \nodata &   \nodata \\
AAT119508   &   1   &   4.12    &   19.86   &   -0.59   &   1.62    &   \nodata &   \nodata &   \nodata \\
AAT120336   &   1   &   7.16    &   19.67   &   -0.63   &   1.59    &   452 &   \nodata &   \nodata \\
AAT120976   &   1   &   7.56    &   19.99   &   -1.27   &   1.29    &   595 &   \nodata &   \nodata \\
C104    &   1   &   8.96    &   19.40   &   -1.54   &   1.19    &   448 &   \nodata &   \nodata \\
C115    &   1   &   12.30   &   19.51   &   -1.62   &   1.16    &   \nodata &   \nodata &   \nodata \\
C123    &   1   &   10.46   &   20.24   &   -0.98   &   1.42    &   \nodata &   \nodata &   \nodata \\
C130    &   1   &   10.92   &   19.85   &   -1.63   &   1.15    &   \nodata &   \nodata &   \nodata \\
C133    &   1   &   3.37    &   19.30   &   -1.00   &   1.40    &   \nodata &   \nodata &   \nodata \\
C138    &   1   &   3.08    &   19.97   &   -1.16   &   1.34    &   \nodata &   \nodata &   \nodata  \\
C140    &   1   &   3.25    &   19.83   &   -1.13   &   1.35    &   \nodata &   \nodata &   \nodata \\
C146    &   1   &   4.45    &   19.94   &   -0.70   &   1.56    &   \nodata &   \nodata &   \nodata \\
C147    &   1   &   2.98    &   20.06   &   -1.24   &   1.31    &   \nodata &   \nodata &   \nodata \\
C149    &   1   &   6.85    &   19.68   &   -1.70   &   1.13    &   389 &   \nodata &   \nodata \\
C150    &   1   &   2.24    &   19.78   &   -1.00   &   1.40    &   \nodata &   \nodata &   \nodata \\
C154    &   1   &   2.97    &   19.58   &   -1.00   &   1.40    &   \nodata &   \nodata &   \nodata \\
C157    &   1   &   3.38    &   19.21   &   -1.00   &   1.40    &   \nodata &   \nodata &   \nodata \\
C159    &   1   &   4.44    &   19.93   &   -0.34   &   1.77    &   \nodata &   \nodata &   \nodata \\
C160    &   1   &   3.33    &   19.99   &   -1.00   &   1.40    &   \nodata &   \nodata &   \nodata \\
C164    &   1   &   3.79    &   19.60   &   -1.00   &   1.40    &   \nodata &   \nodata &   \nodata \\
C167    &   1   &   7.22    &   20.41   &   -1.00   &   1.40    &   \nodata &   \nodata &   \nodata \\
F2GC14  &   1   &   6.97    &   20.82   &   -0.89   &   1.46    &   \nodata &   \nodata &   \nodata \\
F2GC31  &   1   &   7.25    &   20.15   &   -1.56   &   1.18    &   \nodata &   \nodata &   \nodata \\
PFF034  &   1   &   12.62   &   19.38   &   -1.29   &   1.28    &   605 &   \nodata &   \nodata \\
PFF041  &   1   &   4.00    &   19.01   &   -1.37   &   1.25    &   456 &   \nodata &   \nodata \\
PFF052  &   1   &   4.58    &   19.42   &   -1.44   &   1.22    &   462 &   \nodata &   \nodata \\
PFF059  &   1   &   4.00    &   19.48   &   -0.41   &   1.73    &   525 &   \nodata &   \nodata \\
PFF063  &   1   &   8.17    &   19.45   &   -2.19   &   0.96    &   554 &   \nodata &   \nodata \\
PFF066  &   1   &   6.79    &   19.44   &   -1.00   &   1.41    &   530 &   -0.81$\pm$0.18  & 12: \\
PFF083  &   1   &   7.61    &   19.44   &   -0.88   &   1.47    &   458 &   \nodata &   \nodata \\
R203    &   1   &   8.32    &   19.38   &   -2.00   &   1.02    &   455 &   -1.02$\pm$0.12  &   6.1$\pm$1.1   \\
C113    &   1   &   19.68   &   19.12   &   -1.61   &   1.16    &   \nodata &   \nodata &   \nodata \\
C132    &   1   &   9.64    &   18.90   &   -1.34   &   1.26    &   436 &   \nodata &   \nodata & \nodata & \nodata \\
C137    &   1   &   2.67    &   19.04   &   -1.00   &   1.40    &   \nodata &   \nodata &   \nodata \\
PFF016  &   1   &   12.56   &   19.33   &   -0.56   &   1.64    &   505 &   -0.35$\pm$0.38  &    10.5: \\
PFF021  &   1   &   9.71    &   18.81   &   -1.77   &   1.10    &   594 &   \nodata &   \nodata \\
PFF023  &   1   &   15.02   &   18.97   &   -1.27   &   1.29    &   457 &   -1.06$\pm$0.13  &    7.0$\pm$1.2 \\
PFF031  &   1   &   12.35   &   18.95   &   -1.35   &   1.26    &   444 &   \nodata &   \nodata \\
PFF035  &   1   &   11.66   &   19.13   &   -0.32   &   1.78    &   627 &   \nodata &   \nodata \\
C043    &   1   &   10.47   &   18.07   &   -1.24   &   1.30    &   518 &   -1.21$\pm$0.04  &    12.0$\pm$2.2 \\
C135    &   1   &   2.75    &   18.90   &   -1.00   &   1.40    &   \nodata &   \nodata &   \nodata \\
C153    &   1   &   2.93    &   18.23   &   -1.00   &   1.40    &   \nodata &   \nodata &   \nodata \\
G221    &   1   &   9.41    &   18.83   &   -0.82   &   1.50    &   390 &   \nodata &   \nodata \\
G293    &   1   &   9.50    &   18.69   &   -1.69   &   1.13    &   581 &   -1.29$\pm$0.07  &    12.0$\pm$3.6 \\
PFF011  &   1   &   22.51   &   18.55   &   -1.55   &   1.18    &   616 &   \nodata &   \nodata & &  \\
AAT118198   &   2   &   5.46    &   19.03   &   -0.17   &   1.89    &   575 &   \nodata &   \nodata & 27.94 for G2 & 121.58 for G2\\
C006    &   2   &   2.10    &   16.51   &   -0.55   &   1.64    &   \nodata &   -0.51:  &    11.00: \\
C018    &   2   &   4.95    &   16.89   &   -1.05   &   1.39    &   480 &   -0.71$\pm$0.10  &    8.0$\pm$1.6  \\
C030    &   2   &   10.87   &   16.68   &   -0.67   &   1.57    &   778 &   -0.27$\pm$0.05  &    6.6$\pm$1.4  \\
C032    &   2   &   12.47   &   17.85   &   -0.31   &   1.79    &   718 &   -0.10$\pm$0.20  &    13: \\
C037    &   2   &   11.95   &   17.96   &   -0.86   &   1.47    &   612 &   -0.53$\pm$0.19  &    12.0$\pm$4.0 \\
C142    &   2   &   1.85    &   17.64   &   -1.00   &   1.40    &   \nodata &   \nodata &   \nodata \\
C145    &   2   &   3.56    &   17.81   &   -1.27   &   1.29    &   \nodata &   \nodata &   \nodata \\
C152    &   2   &   2.75    &   17.82   &   -1.00   &   1.40    &   \nodata &   \nodata &   \nodata \\
F2GC69  &   2   &   10.51   &   19.30   &   -0.63   &   1.60    &   \nodata &   \nodata &   \nodata \\
G284    &   2   &   5.93    &   19.41   &   -0.56   &   1.64    &   479 &   \nodata &   \nodata  \\
K131    &   2   &   3.84    &   18.74   &   -0.18   &   1.89    &   639 &   \nodata &   \nodata \\
PFF079  &   2   &   10.72   &   19.14   &   -1.33   &   1.27    &   410 &   \nodata &   \nodata \\
R223    &   2   &   6.60    &   18.19   &   -0.83   &   1.49    &   776 &   \nodata &   \nodata \\
C117    &   2   &   9.73    &   19.26   &   -0.30   &   1.79    &   \nodata &   \nodata &   \nodata \\
C168    &   2   &   7.91    &   19.71   &   -1.00   &   1.40    &   \nodata &   \nodata &   \nodata \\
F1GC14  &   2   &   12.11   &   19.67   &   -1.49   &   1.20    &   \nodata &   \nodata &   \nodata \\
PFF029  &   2   &   8.87    &   19.10   &   -4.20   &   0.40    &   570 &   \nodata &   \nodata \\
C161    &   2   &   7.28    &   19.26   &   -0.39   &   1.74    &   425 &   \nodata &   \nodata \\
C003    &   2   &   8.10    &   17.08   &   -0.41   &   1.72    &   562 &   -0.13$\pm$0.05  &    5.8$\pm$1.2  \\
C004    &   2   &   10.53   &   17.50   &   -1.42   &   1.23    &   689 &   -1.44$\pm$0.12  &    10: \\
C007    &   2   &   9.18    &   16.64   &   -1.21   &   1.32    &   595 &   -0.93$\pm$0.09  &    12.0$\pm$4.0 \\
C012    &   2   &   11.26   &   17.36   &   -0.34   &   1.77    &   \nodata &   \nodata &   \nodata \\
C014    &   2   &   18.31   &   17.41   &   -0.94   &   1.44    &   705 &   -0.85$\pm$0.16  &    10.0$\pm$4.0 \\
C019    &   2   &   7.59    &   17.55   &   -0.93   &   1.44    &   632 &   -0.91$\pm$0.14  &    12: \\
C025    &   2   &   8.49    &   17.96   &   -0.39   &   1.74    &   703 &   -0.21$\pm$0.07  &    8.0$\pm$2.6  \\
C029    &   2   &   21.02   &   17.53   &   -0.44   &   1.71    &   726 &   -0.17$\pm$0.05  &    8.0$\pm$1.8  \\
C036    &   2   &   12.95   &   17.94   &   -1.61   &   1.16    &   703 &   -1.07$\pm$0.12  &    12: \\
C139    &   2   &   2.77    &   18.86   &   -0.53   &   1.65    &   \nodata &   \nodata &   \nodata &  &  \\
C151    &   2   &   3.26    &   19.95   &   0.05    &   2.10    &   \nodata &   \nodata &   \nodata \\
C165    &   2   &   5.27    &   18.17   &   -0.42   &   1.71    &   \nodata &   \nodata &   \nodata \\
G170    &   2   &   8.86    &   18.73   &   -0.57   &   1.63    &   636 &   -0.26$\pm$0.18  &   12: \\
WHH09   &   2   &   4.35    &   18.30   &   -0.35   &   1.76    &   315 &   \nodata &   \nodata \\
WHH16   &   2   &   3.19    &   18.60   &   -0.27   &   1.82    &   661 &   -0.20$\pm$0.04  &   8.0$\pm$0.02   \\
WHH22   &   2   &   5.04    &   18.03   &   -0.99   &   1.41    &   \nodata &   \nodata &   \nodata \\
C126    &   3   &   10.45   &   22.66   &   -1.67   &   1.14    &   \nodata &   \nodata &   \nodata&\nodata&\nodata \\
AAT117287   &   3   &   3.53    &   20.45   &   -1.62   &   1.16    &   554 &   \nodata &   \nodata \\
C111    &   3   &   21.69   &   21.36   &   -2.20   &   0.96    &   \nodata &   \nodata &   \nodata \\
C114    &   3   &   12.67   &   21.42   &   -0.27   &   1.82    &   \nodata &   \nodata &   \nodata \\
C118    &   3   &   10.38   &   20.67   &   -0.44   &   1.71    &   \nodata &   \nodata &   \nodata &  &  \\
C124    &   3   &   11.99   &   21.57   &   -1.30   &   1.28    &   \nodata &   \nodata &   \nodata \\
C125    &   3   &   12.34   &   20.66   &   -0.91   &   1.45    &   \nodata &   \nodata &   \nodata \\
C127    &   3   &   9.92    &   21.62   &   -1.08   &   1.38    &   \nodata &   \nodata &   \nodata \\
C131    &   3   &   11.85   &   20.20   &   -1.65   &   1.15    &   \nodata &   \nodata &   \nodata \\
C136    &   3   &   3.84    &   21.19   &   -1.84   &   1.08    &   \nodata &   \nodata &   \nodata \\
C143    &   3   &   2.45    &   20.52   &   -1.70   &   1.13    &   \nodata &   \nodata &   \nodata \\
C158    &   3   &   5.39    &   20.06   &   -1.05   &   1.39    &   \nodata &   \nodata &   \nodata \\
C163    &   3   &   4.05    &   20.43   &   -1.18   &   1.33    &   \nodata &   \nodata &   \nodata \\
C170    &   3   &   7.93    &   22.16   &   -2.27   &   0.93    &   \nodata &   \nodata &   \nodata \\
C172    &   3   &   10.83   &   20.91   &   -1.99   &   1.02    &   \nodata &   \nodata &   \nodata \\
C174    &   3   &   9.17    &   21.42   &   -0.63   &   1.60    &   \nodata &   \nodata &   \nodata \\
C176    &   3   &   8.41    &   21.30   &   -2.58   &   0.84    &   \nodata &   \nodata &   \nodata \\
C179    &   3   &   9.91    &   21.04   &   -2.47   &   0.87    &   \nodata &   \nodata &   \nodata \\
F2GC18  &   3   &   6.93    &   21.15   &   -1.00   &   1.40    &   \nodata &   \nodata &   \nodata \\
F2GC20  &   3   &   6.59    &   21.01   &   -0.53   &   1.65    &   \nodata &   \nodata &   \nodata \\
F2GC28  &   3   &   6.39    &   21.16   &   -0.84   &   1.48    &   \nodata &   \nodata &   \nodata \\
C105    &   3   &   9.73    &   21.76   &   -0.44   &   1.70    &   \nodata &   \nodata &   \nodata &  &  \\
C112    &   3   &   22.46   &   21.31   &   -0.91   &   1.45    &   \nodata &   \nodata &   \nodata \\
C119    &   3   &   8.45    &   20.67   &   -4.77   &   0.27    &   \nodata &   \nodata &   \nodata \\
C120    &   3   &   11.67   &   21.43   &   -0.60   &   1.61    &   \nodata &   \nodata &   \nodata \\
C121    &   3   &   9.91    &   22.45   &   -2.49   &   0.86    &   \nodata &   \nodata &   \nodata \\
C122    &   3   &   9.84    &   22.33   &   -1.00   &   1.40    &   \nodata &   \nodata &   \nodata \\
C128    &   3   &   9.74    &   20.95   &   -0.14   &   1.91    &   \nodata &   \nodata &   \nodata \\
C129    &   3   &   12.53   &   20.83   &   -0.73   &   1.54    &   \nodata &   \nodata &   \nodata \\
C134    &   3   &   3.29    &   20.71   &   -0.78   &   1.52    &   \nodata &   \nodata &   \nodata \\
C141    &   3   &   2.68    &   20.94   &   -1.86   &   1.07    &   \nodata &   \nodata &   \nodata \\
C144    &   3   &   3.87    &   21.96   &   -1.95   &   1.04    &   \nodata &   \nodata &   \nodata \\
C148    &   3   &   5.17    &   20.21   &   -2.62   &   0.82    &   \nodata &   \nodata &   \nodata \\
C155    &   3   &   7.85    &   21.36   &   -1.00   &   1.41    &   \nodata &   \nodata &   \nodata \\
C162    &   3   &   5.04    &   20.91   &   -2.60   &   0.83    &   \nodata &   \nodata &   \nodata \\
C166    &   3   &   4.97    &   20.72   &   -1.00   &   1.40    &   \nodata &   \nodata &   \nodata \\
C171    &   3   &   8.64    &   20.67   &   -1.11   &   1.36    &   \nodata &   \nodata &   \nodata \\
C173    &   3   &   9.35    &   21.06   &   -0.19   &   1.88    &   \nodata &   \nodata &   \nodata \\
C175    &   3   &   7.35    &   21.84   &   -0.91   &   1.45    &   \nodata &   \nodata &   \nodata &  &  \\
C178    &   3   &   8.78    &   21.53   &   -1.34   &   1.26    &   \nodata &   \nodata &   \nodata \\
F1GC20  &   3   &   9.95    &   21.22   &   -0.58   &   1.62    &   \nodata &   \nodata &   \nodata \\
F1GC21  &   3   &   11.01   &   21.36   &   -0.22   &   1.85    &   \nodata &   \nodata &   \nodata \\
F1GC34  &   3   &   12.16   &   20.96   &   -0.48   &   1.68    &   \nodata &   \nodata &   \nodata \\
F2GC70  &   3   &   9.36    &   20.04   &   -1.96   &   1.04    &   \nodata &   \nodata &   \nodata \\
C116    &   3   &   11.72   &   21.84   &   -0.37   &   1.75    &   \nodata &   \nodata &   \nodata \\
C156\tablenotemark{b}   &   *   &   4.83    &   17.65   &   -0.18   &   1.89    &   \nodata &   \nodata &   \nodata \\
C169\tablenotemark{b}   &   *   &   7.74    &   20.39   &   -2.08   &   1.00    &   \nodata &   \nodata &   \nodata  \\
F1GC15\tablenotemark{b} &   *   &   12.11   &   19.53   &   -0.04   &   2.01    &   \nodata &   \nodata &   \nodata \\
\enddata
\tablecomments {Values in column 1 and in columns 3 to 6 are from
McLaughlin et al .(2008), column 2 represents the group membership
found by CA,values in column 7 are from Woodley et
al. (2007), values in columns 8 to 11 have been derived in the
present work.}\\
 \tablecomments{':'s are used for the errors of
[Z/H]s and Ages when error of [Z/H] $
>0.4$ dex and error of Age $> 4$ Gyr} \tablenotetext{b}{These GCs
have not been considered during the CA as they are outliers.}
\end{deluxetable}

\clearpage

\begin{deluxetable}{ccccccc}
\tabletypesize{\scriptsize}
 \tablewidth{0pt} \tablecaption{Result
of  PCA analysis}\tablehead{
\colhead{Set} & \colhead{components} & \colhead{\% of} & \colhead{no. of} & \multicolumn{2}{c}{eigen} & \colhead{significant} \\
\colhead{} & \colhead{} & \colhead{variations} & \colhead{significant} & \multicolumn{2}{c}{vectors} & \colhead{parameters} \\
\colhead{} & \colhead{} & \colhead{} & \colhead{eigen} & \colhead{} & \colhead{} & \colhead{} \\
\colhead{} & \colhead{} & \colhead{} & \colhead{vectors} & \colhead{} & \colhead{} & \colhead{} \\
\colhead{} & \colhead{} & \colhead{} & \colhead{corresponding} & \colhead{} & \colhead{} & \colhead{} \\
\colhead{} & \colhead{} & \colhead{} & \colhead{to variation} & \colhead{} & \colhead{} & \colhead{} \\
\colhead{} & \colhead{} & \colhead{} & \colhead{$>$ 90\%} & \colhead{(1)} & \colhead{(2)} & \colhead{} \\
} \startdata
S1([Fe/H], & 1 & 67.54 & 2 & (0.0164, & (-0.0233, & T$_{1}$, \\
T$_{1}$,c,$\mu_{0}$, & 2* & 91.95* &  & -0.1712, & -0.0121, & $\mu_{0}$, \\
W$_{0}$,R$_{c}$, & 3 & 97.95 &  & 0.0396, & -0.0006, & r$_{h}$, \\
r$_{h}$,$\sigma_{p,0}$, & 4 & 98.90 &  & -0.2641, & 0.0185, & $\sigma_{p,0}$, \\
R$_{gc}$)& 5 & 99.50 &  & 0.0638, & 0.0012, & R$_{gc}$ \\
 & 6 & 99.90 &  & -0.0906, & 0.0188, &  \\
 & 7 & 99.97 &  & -0.1053, & 0.0526, &  \\
 & 8 & 99.99 &  & 0.9348, & 0.0587, &  \\
 & 9 & 100.00 &  & -0.0446) & 0.9962) &  \\ \tableline
S2($\mu_{0}$,r$_{h}$, & 1 & 70.44 & 2 & (-0.2960, & (0.0179, & $\mu_{0}$, \\
$\sigma_{p,0}$,R$_{gc}$) & 2* & 93.29* &  & -0.1281, & 0.0945, & R$_{h}$, \\
 & 3 & 99.38 &  & 0.9430, & 0.0933, & $\sigma_{p,0}$, \\
 & 4 & 100.00 &  & -0.0714) & 0.9909) & R$_{gc}$ \\ \tableline
S3($\mu_{0}$,r$_{h}$, & 1* & 90.39* & 1 & (-0.2924, &  & $\mu_{0}$, \\
$\sigma_{p,0}$,c) & 2 & 98.93 &  & -0.1174, &  & R$_{h}$, \\
 & 3 & 99.89 &  & 0.947, &  & $\sigma_{p,0}$, \\
 & 4 & 100.00 &  & 0.100) &  & c \\ \tableline
S4($\mu_{0}$,r$_{h}$, & 1 & 91.35* & 1 & (-0.3003, &  & $\mu_{0}$, \\
$\sigma_{p,0}$) & 2 & 99.23 &  & -0.1291, &  & R$_{h}$, \\
 & 3 & 100.00 &  & 0.9450) &  & $\sigma_{p,0}$ \\ \tableline
S5([Fe/H], & 1 & 71.72 & 2 & (-0.135, & (0.957, & [Fe/H], \\
c,R$_{c}$) & 2* & 91.6* &  & -0.488, & 0.160, & c, \\
 & 3 & 100.00 &  & 0.861) & 0.240) & R$_{c}$ \\ \tableline
S6($\mu_{0}$,r$_{h}$, & 1 & 88.99* & 1 & (-0.2821, &  & $\mu_{0}$, \\
$\sigma_{p,0}$,W$_{0}$) & 2* & 98.15 &  & -0.1130, &  & R$_{h}$, \\
 & 3 & 99.78 &  & 0.9486, &  & $\sigma_{p,0}$, \\
 & 4 & 100.00 &  & 0.100) &  & W$_{0}$\\ \tableline
S7($<\mu_v>_h,r_h,\sigma_{p,0}$)& 1& 91.78*& 1& (-0.238,& &
$<\mu_v>_h$,\\
 & 2& 99.57& & -0.149, & &$r_h$,\\
 & 3& 100.00& & 0.9600)& & $\sigma_{p,0}$\\ \tableline
 S8($<\mu_v>_h,r_h,\sigma_{p,0}$& 1& 88.51& 2&(-0.216,&
 (0.333,& $<\mu_v>_h$,\\
 $,W_0$) & 2& 98.39*& & -0.110,& 0.926,& $r_h$,\\
  & 3& 99.62& & -0.967,& 0.178,& $\sigma_{p,0}$,\\
  & 4& 100.00& &0.079)&  0.031)& $W_0$\\ \tableline
  S9($<\mu_v>_h, r_h, \sigma_{p,0},c$)&1& 90.13*& 1& (-0.222,&
  &$<\mu_v>_h$,\\
  & 2& 99.14& & -0.120,& & $r_h$,\\
  & 3& 99.59& & 0.966,& & $\sigma_{p,0}$,\\
  & 4& 100.00&  &0.056)&   & c \\
\enddata
\end{deluxetable}

\clearpage

\begin{deluxetable}{cccc}
\tabletypesize{\scriptsize}
 \tablewidth{0pt} \tablecaption{Mean
values of the observed and derived parameters in three groups of
GCs found by CA} \tablehead{ \colhead{} & \colhead{G1} &
\colhead{G2} & \colhead{G3} } \startdata
No & 47 & 35 & 45 \\
Age (Gyr) & 10.202$\pm0.845$ & 9.447$\pm0.814$ & \nodata \\
$\left[Z/H\right]$ (dex)& -0.985$\pm0.152$ & -0.553$\pm0.105$ & \nodata \\
$\mu_{0}$ $(mag~arcsec^{-2})$ & 15.828$\pm0.127$ & 13.414$\pm0.412$ & 19.077$\pm0.170$ \\
log$R_{c}$ (pc)& -0.011$\pm$0.022 & -0.429$\pm$0.142 & 0.363$\pm$0.055 \\
log$R_{tid}$ (pc)& 2.284$\pm$0.039 & 2.432$\pm$0.082 & 2.244$\pm$0.064 \\
log$r_{h}$ (pc)& 0.523$\pm$0.0162 & 0.571$\pm$0.043 & 0.786$\pm$0.027 \\
log$M_{tot}$ $(M_{\odot})$ & 5.642$\pm$0.031 & 6.187$\pm$0.062 & 5.023$\pm$0.042 \\
$\rho_{0}$ $(M_{\odot}pc^{-3})$ & (2.605$\pm0.039$)$\times$10$^{4}$ & (6.773$\pm6.54$)$\times$10$^{9}$ & (2.870$\pm2.79$)$\times$10$^{3}$ \\
$\sigma_{p,0}$ (km/sec) & 9.984$\pm$0.400 & 20.528$\pm1.130$ & 3.533$\pm0.144$ \\
$t_{rh}$ ($10^{9}$ years)& 1.502$\pm$0.099 & 4.762$\pm$0.088 & 2.538$\pm$0.309 \\
R$_{gc}$ (kpc)& 7.532$\pm0.660$ & 7.992$\pm0.745$ & 8.938$\pm0.613$ \\
T$_{1}$ (mag) & 19.462$\pm0.081$ & 18.218$\pm0.159$& 21.156$\pm0.092$ \\
$[Fe/H]$ (dex) & -1.172$\pm0.068$ & -0.816$\pm0.122$ & -1.317$\pm0.134$ \\
$(C-T_{1})_{0}$ (mag) & 1.349$\pm$0.030 & 1.539$\pm$0.050 & 1.322$\pm$0.052 \\
$c$&2.255$\pm$0.053& 2.811$\pm$ 0.142& 1.784$\pm$0.106\\
$W_{0}$ & 6.749$\pm$0.088 & 8.246$\pm$0.539 & 5.662$\pm$0.316 \\
$<\mu_v>_h$ $(mag ~arcsec^{-2})$& 17.722$\pm$0.114& 16.778$\pm$0.185&20.640$\pm$0.136\\
$V_r$ (km/sec)& 520.5$\pm$16.5& 608.6$\pm$26.3 & \nodata \\
e (Ellipticity)& 0.068$\pm$0.05& 0.157$\pm$0.14& 0.068$\pm$0.04\\
V$_{rot}$ (km/sec) & 43.96 & 27.94 & \nodata \\
$\theta_{0}$ $(deg)$ & 185.83(major) & 285.12(minor) & \nodata \\
$\sigma_{v}$ (km/sec) & 102.082 & 121.578 & \nodata \\
x & 0.43 & 0.23 & \nodata \\
$\Lambda$=0.3x & 0.129 & 0.069 & \nodata \\
\enddata
\end{deluxetable}

\end{document}